\documentclass[lettersize,journal]{IEEEtran}
\usepackage{amsmath,amsfonts}
\usepackage{algorithmic}
\usepackage{algorithm}
\usepackage{array}
\usepackage{textcomp}
\usepackage{stfloats}
\usepackage{url}
\usepackage{verbatim}
\usepackage{graphicx}
\usepackage{cite}
\usepackage{wasysym}
\usepackage{fontawesome}
\usepackage{adjustbox}
\usepackage{indentfirst}
\usepackage{multirow}
\usepackage{graphicx}
\usepackage{color}
\usepackage{xcolor}
\usepackage{colortbl}
\usepackage{listings}
\usepackage{enumerate}
\usepackage{amsmath,amsthm,amscd,amssymb,latexsym,upref,stmaryrd}
\usepackage{amsfonts,epsfig}
\usepackage{CJK}
\usepackage{float}
\usepackage{cite}
\usepackage{lipsum}
\usepackage{multicol}
\usepackage{setspace}
\usepackage{bm}
\usepackage{amsthm}
\usepackage{amsmath}
\usepackage{amssymb}
\usepackage{amsfonts}
\usepackage{arydshln}
\usepackage{stfloats}
\usepackage{booktabs}
\usepackage{threeparttable}
\usepackage{stackengine}
\newtheorem{remark}{Remark}
\usepackage[utf8]{inputenc}
\usepackage{fancyhdr}

\usepackage[colorlinks,linkcolor=black,anchorcolor=black,citecolor=black,urlcolor=black]{hyperref}
\usepackage{makecell}

\newcommand{\xrowht}[2][0]{\addstackgap[.5\dimexpr#2\relax]{\vphantom{#1}}}

\usepackage{subcaption}

\newcolumntype{P}[1]{>{\centering\arraybackslash}m{#1}}

\begin{document}

\title{Physics-Informed Implicit Neural Representation for Wireless Imaging in RIS-Aided ISAC System}

\author{Yixuan~Huang,~\IEEEmembership{Graduate Student Member, IEEE,} Jie~Yang,~\IEEEmembership{Member, IEEE,} Chao-Kai~Wen,~\IEEEmembership{Fellow,~IEEE,} and Shi~Jin,~\IEEEmembership{Fellow,~IEEE}
\thanks{Yixuan Huang is with the School of Information Science and Engineering, Southeast University, Nanjing 210096, China (e-mail: huangyx@seu.edu.cn).

Jie Yang is with the Key Laboratory of Measurement and Control of Complex Systems of Engineering, Ministry of Education, and the Frontiers Science Center for Mobile Information Communication and Security, Southeast University, Nanjing 210096, China (e-mail: yangjie@seu.edu.cn).

Chao-Kai Wen is with the Institute of Communications Engineering, National Sun Yat-sen University, Kaohsiung 80424, Taiwan. (e-mail: chaokai.wen@mail.nsysu.edu.tw).

Shi Jin is with the School of Information Science and Engineering, and the Frontiers Science Center for Mobile Information Communication and Security, Southeast University, Nanjing 210096, China (e-mail: jinshi@seu.edu.cn).

Digital Object Identifier 10.1109/TWC***
}
}

\maketitle

\begin{abstract}

Wireless imaging has become a vital function in future integrated sensing and communication (ISAC) systems.
However, traditional model-based and data-driven deep learning imaging methods face challenges related to multipath extraction, dataset acquisition, and multi-scenario adaptation.
To overcome these limitations, this study innovatively combines implicit neural representation (INR) with explicit physical models to realize wireless imaging in reconfigurable intelligent surface (RIS)-aided ISAC systems.
INR employs neural networks (NNs) to project physical locations to voxel values, which is indirectly supervised by measurements of channel state information with physics-informed loss functions.
The continuous shape and scattering characteristics of targets are embedded into NN parameters through training, enabling arbitrary image resolutions and off-grid voxel value prediction.
Additionally, three issues related to INR-based imager are further addressed.
First, INR is generalized to enable efficient imaging under multipath interference by jointly learning image and multipath information.
Second, the imaging speed and accuracy for dynamic targets are enhanced by embedding prior image information.
Third, imaging results are employed to assist in RIS phase design for improved communication performance.
Extensive simulations demonstrate that the proposed INR-based imager significantly outperforms traditional model-based methods with super-resolution abilities, and the focal length characteristics of the imaging system is revealed.
Moreover, communication performance can benefit from the imaging results.
Part of the source code for this paper can be accessed at \url{https://github.com/kiwi1944/INRImager}

\end{abstract}


\section{Introduction}
Integrated sensing and communication (ISAC) has been incorporated as one of the primary application scenarios for sixth-generation communication systems \cite{itu2023framework,wei2023integrated}.  
Various sensing functions, including recognition, localization, and mapping, are anticipated to be integrated into future communication systems \cite{lei2024deep,lin2024environment,wei20225g}.  
Rather than being confined to detecting and locating point targets, ISAC has shown a tendency toward sensing fine-grained shape and scattering characteristics of extended targets through wireless imaging \cite{wang2024cramer,zheng2024random,lyu2024compressed}.  
It offers notable benefits in all-day sensing and privacy protection without the use of visible light, inspiring numerous applications such as health monitoring and low-altitude surveillance \cite{hu2022metasketch,huang2025cooperative}.  

Wireless imaging in traditional communication systems has been hindered by the need for large antenna arrays and short imaging distances \cite{sheen2001three}.
Recently, reconfigurable intelligent surfaces (RISs) have been recognized as a key enabler for ISAC to partially address these issues by employing large, low-cost, and energy-efficient arrays in the vast communication environment \cite{tang2021path,li2024radio,wang2025dreamer}.
Additionally, RIS phase variations can manually adjust the electromagnetic field to provide diverse sensing perspectives and channel state information (CSI) measurements \cite{he2022high}.  
Given these advantages, we aim to realize extended target imaging within RIS-aided ISAC systems.  

\begin{figure}
    \centering
    \includegraphics[width=0.95\linewidth]{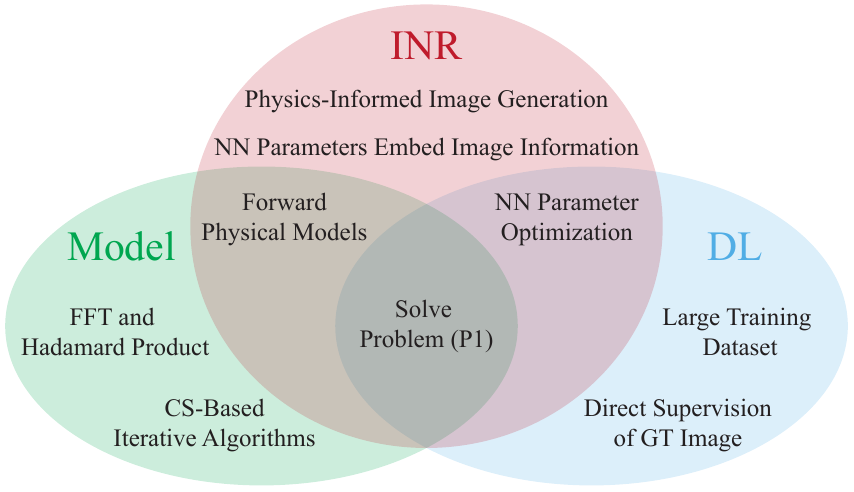}
    \captionsetup{font=footnotesize}
     \caption{Relationship between model-based, data-driven DL, and INR methods.} 
    \label{fig-introduction-compare}
\end{figure}

Wireless imaging extracts target information from CSI measurements.  
Generally, the measurement model can be expressed as  
\begin{equation}\label{eq-general-model}
\mathbf{y} = f(\boldsymbol{\sigma}) + \mathbf{n},
\end{equation}  
where $\mathbf{y}\in\mathbb{C}^{N_{\text{m}}}$ is the CSI measurements, and $N_{\text{m}}$ denotes the measurement number.  
The scattering coefficient image of the region of interest (ROI) is denoted as $\boldsymbol{\sigma}\in\mathbb{R}^{N_{\text{v}}}$, where $N_{\text{v}}$ is the number of voxels.  
$f(\cdot)$ is the forward model describing the relationship between $\boldsymbol{\sigma}$ and CSI, which varies with system configurations.  
$\mathbf{n}\in\mathbb{C}^{N_{\text{m}}}$ represents measurement noise.  
Various imaging techniques aim to solve the inverse problem of \eqref{eq-general-model} by optimizing the image $\boldsymbol{\sigma}$, formulated as  
\begin{equation*}
\text{(P1)}\quad \boldsymbol{\sigma}^\star = \mathop{\arg\!\min}\limits_{\boldsymbol{\sigma}}~ \mu(f(\boldsymbol{\sigma}), \mathbf{y}) + \rho(\boldsymbol{\sigma}),
\end{equation*}
where $\mu(f(\boldsymbol{\sigma}), \mathbf{y})$ measures the error between $f(\boldsymbol{\sigma})$ and $\mathbf{y}$, and $\rho(\boldsymbol{\sigma})$ is the regularization term capturing the image prior.  

In traditional model-based imaging methods, Fourier transform (FT) and compressed sensing (CS) both require the knowledge of the forward model $f(\cdot)$.
Specifically, FT implements $f(\cdot)$ as a spatial integral, whereas $f(\cdot)$ should be discretized into a high-dimensional sensing matrix in CS theory \cite{tong2025computational,huang2024ris}.
According to \cite{huang2024fourier}, FT- and CS-based methods exhibit distinct advantages in terms of computational complexity, imaging accuracy, and measurement requirements.
However, both approaches require multipath extraction to preserve only the transmitter (TX)-ROI-RIS-receiver (RX) path \cite{huang2024fourier,he2022high,hu2022metasketch}, potentially discarding target information involved in other multipath components.
Moreover, errors in modeling $f(\cdot)$ may significantly distort the imaging results \cite{tong2025computational,huang2025cooperative}.

In contrast, data-driven deep learning (DL) methods reduce reliance on $f(\cdot)$ by learning the relationship between $\mathbf{y}$ and $\boldsymbol{\sigma}$ from a large dataset containing numerous $(\mathbf{y}, \boldsymbol{\sigma})$ pairs \cite{lu2024deep}.  
Although DL techniques have demonstrated superior imaging performance, constructing such a large dataset is challenging since the ground-truth (GT) ROI image is difficult to obtain \cite{li2020intelligent}.  
Furthermore, transferring well-trained model parameters to new imaging environments, targets, and resolutions is complicated \cite{qi2023resource}, requesting retraining the NNs.
This has limited the application of data-driven DL methods.
While researchers have attempted to integrate physical models into learning \cite{guo2023physics,huang2025integrated}, the above challenges in model-based and data-driven DL methods remain inadequately addressed.  
 
Recently, implicit neural representation (INR) has shown excellent performance in optical 3D reconstruction \cite{mildenhall2021nerf} and has been applied to communication systems for radio map generation \cite{zhao2023nerf2}, data suppression \cite{wu2024embracing}, and RIS phase design \cite{yang2024codebook}.
INR leverages a neural network (NN) to learn a continuous function that maps input coordinates to their corresponding signal values, thereby encoding the signal or scene within NN parameters.
Specifically, the term \textit{implicit neural representation} can be decomposed into:  
1) \textit{implicit representation}, indicating that INR uses implicit and continuous functions to represent sensing results rather than explicit and discretized forms such as images and point clouds, and
2) \textit{neural representation}, meaning that these implicit functions are realized by NNs.

Inspired by the great success of INR in the above fields, this study extends its application to wireless imaging by implicitly extracting and representing radio images using NNs.
Consequently, the image optimization problem (P1) is transformed into the optimization of NN parameters, which can be executed using DL techniques and accelerated by common GPU resources in future communication systems \cite{boccuzzi2025gpu}.

Compared with model-based methods, INR supports arbitrary differentiable forward models and mitigates the need for multipath extraction.
Although DL techniques are employed, INR does not require GT images or large training datasets \cite{lu2024deep}, since it is indirectly supervised by CSI measurements $\mathbf{y}$ using the physical model $f(\cdot)$ \cite{yu2024ai}.
By employing INR, the continuous shape and scattering characteristics of targets can be learned through NN training.
This information is implicitly embedded into NN parameters, allowing INR to generate images of arbitrary resolutions, which is infeasible for traditional methods.
This feature advances the development of super-resolution and off-grid imaging \cite{guo2025unlocking}.
The conceptual relationship between model-based, data-driven DL, and INR-based methods is illustrated in Fig. \ref{fig-introduction-compare}.

Despite these advantages, several additional issues should be addressed: 

\textbf{First}, INR-based imagers still face the challenge of \textit{inaccurate forward models}, especially in complex environments where multipath components, which are difficult to model in $f(\cdot)$, introduce significant interference.
Although such interference can be suppressed through multipath extraction \cite{huang2024fourier,huang2024ris}, the estimation accuracy is not guaranteed.  
Special TX/RX placements and transmission directions are utilized in \cite{hu2022metasketch}, but they may be impractical in communication systems.  
Complex algorithms are employed to design ISAC beamformers in \cite{wang2024cramer,li2023toward}, but the effects of RISs are not considered.  
Moreover, NNs have been used to approximate $f(\cdot)$ \cite{li2020intelligent,zheng2025hybrid}, requiring large training datasets, which counteract the benefits of INR.  
Alternatively, our study proposes to model multipath interference as trainable parameters within INR, enabling the joint learning of the ROI image and multipath components.

\textbf{Second}, since INR encodes image information into NN parameters, it requires \textit{online training} for each target, leading to \textit{delays} in producing imaging results, which is critical for dynamic target imaging.
Although hierarchical sampling \cite{mildenhall2021nerf} and multiresolution hash encoding \cite{muller2022instant} have been developed to accelerate NN training for INR, they significantly increase algorithmic complexity.  
Unlike the location variation of point targets, the dynamics of extended targets typically manifest as changes in shape, suggesting potential temporal correlations across adjacent time frames \cite{wan2025static,huang2024ris}.  
Exploiting these correlations, \cite{shen2022nerp} has achieved high-accuracy medical imaging and image variation capture.
Inspired by this, we propose to utilize prior image information captured in previous instants to accelerate dynamic target imaging in subsequent steps.

\textbf{Third}, while previous studies have focused on enhanced imaging within communication systems \cite{huang2024fourier,huang2024ris,hu2023imaging}, the benefits of imaging for communication remain underexplored.
\cite{luo2023integrated} proposes using wireless depth images to detect user locations and optimize RIS radiation beams.  
Similarly, human body images are reconstructed and input to an NN to detect the desired sensing regions in \cite{li2019intelligent}.  
According to \cite{wu2018intelligent}, the knowledge of the surrounding environment, which can be supplemented by INR-based imaging results, is crucial for communication-oriented RIS phase design, leading to improved spectral efficiency (SE).  
In this study, we respectively address the mechanisms of imaging-augmented communication by leveraging the sensing results for users inside and outside the ROI.

To summarize, our key contributions are as follows:  
We pioneer the application of INR to wireless imaging in RIS-aided ISAC systems, employing DL methods under the supervision of physical models.
Free of dataset construction, image information is directly extracted from CSI measurements and embedded into NN parameters through training, generating high-accuracy images at arbitrary resolutions, which traditional methods cannot achieve.  
In addition, the following issues are also addressed:
\begin{itemize}
\item \textbf{Imaging under Multipath Interference:}  
We first eliminate ROI-irrelevant multipath components through background calibration.
Then, the INR-based imager is generalized to model ROI-related interference as trainable parameters, which is learned during the imaging process.

\item \textbf{Successive Imaging for Dynamic Targets:}  
We embed prior image information from previous time instants into subsequent imaging through INR parameter initialization, thereby accelerating NN training, reducing output delay, and enhancing imaging performance.

\item \textbf{Imaging-Augmented Communication:}  
We leverage INR-based imaging results to complementary environmental information, which supports RIS phase design to enhance the communication performance for users inside or outside the ROI.
\end{itemize}

The remainder of this paper is structured as follows:  
Sec. \ref{sec-system-model} introduces the RIS-aided ISAC system, and Sec. \ref{sec-inr-imaging} presents INR-based wireless imaging method.  
Sec. \ref{sec-enhanced-imager} addresses the issues of multipath interference and dynamic target imaging.  
Sec. \ref{sec-imaging-augmented-communication} explores imaging-augmented communication performance. 
Simulation results and conclusions are provided in Sec. \ref{sec-simulation} and Sec. \ref{sec-conclusion}, respectively.   

\textbf{Notations}---Scalars (e.g., $a$) are denoted in italics, vectors (e.g., $\mathbf{a}$) in bold, and matrices (e.g., $\mathbf{A}$) in bold capital letters. The $\ell_{1}$- and $\ell_{2}$-norms of $\mathbf{a}$ are denoted by $\|\mathbf{a}\|_{1}$ and $\|\mathbf{a}\|_{2}$, respectively. $j = \sqrt{-1}$ is the imaginary unit. $\text{diag}{(\mathbf{a})}$ constructs a diagonal matrix from the elements of $\mathbf{a}$. The transpose and Hermitian operators are $(\cdot)^{\text{T}}$ and $(\cdot)^{\text{H}}$, respectively.

\section{System Model}
\label{sec-system-model}

\begin{figure}[t]
\centering
\captionsetup{font=footnotesize}
\begin{subfigure}[b]{0.69\linewidth}
\centering
\includegraphics[width=\linewidth]{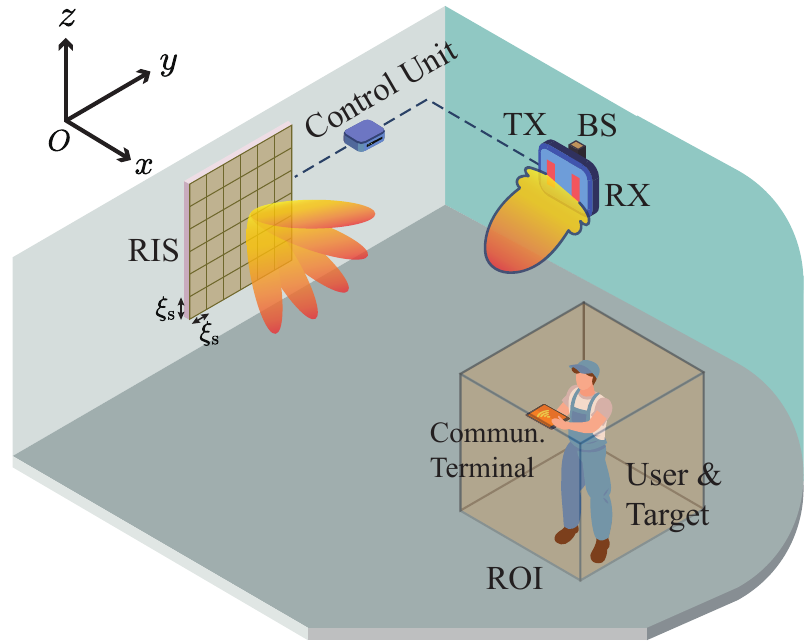}
\caption{User inside the ROI}
\label{fig-model-in}
\end{subfigure}
\begin{subfigure}[b]{0.69\linewidth}
\centering
\includegraphics[width=\linewidth]{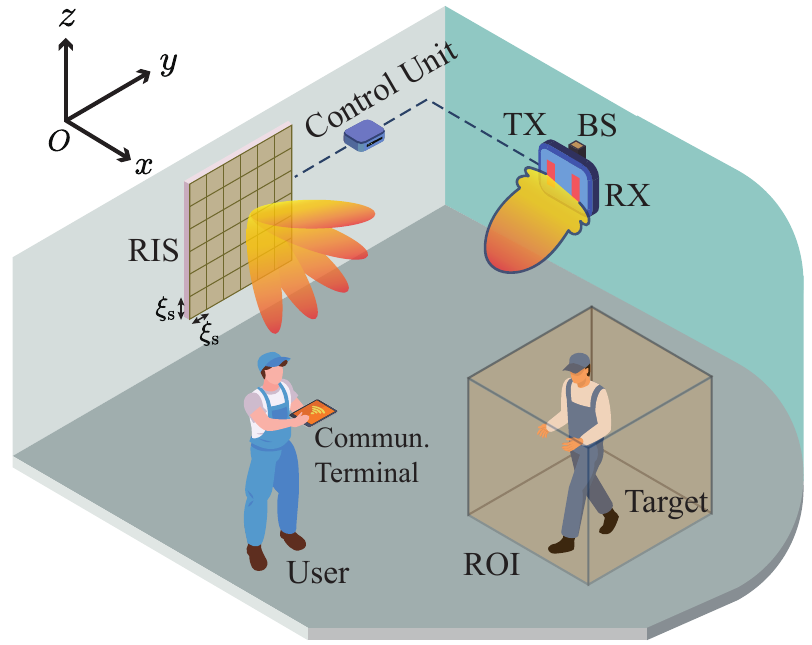}
\caption{User outside the ROI}
\label{fig-model-out}
\end{subfigure}

\caption{Illustration of the proposed RIS-aided ISAC system for wireless imaging.}
\label{fig-model}
\end{figure}

We consider a RIS-aided ISAC system operating in a 3D space $\mathbb{R}^3=\{[x, y, z]^{\text{T}}:x,y,z \in \mathbb{R}\}$, as illustrated in Fig.~\ref{fig-model}.  
A full-duplex BS is employed to enable both user communication and target imaging, using orthogonal frequency division multiplexing (OFDM) signals.  
The BS's TX and RX are equipped with uniform linear arrays (ULAs) consisting of $N_{\text{t}}$ and $N_{\text{r}}$ antennas, respectively, with half-wavelength spacing $\lambda/2$, where $\lambda$ is the wavelength corresponding to the center subcarrier frequency.  
Self-interference at the BS is assumed to be mitigated by sufficiently isolated TX and RX antenna arrays \cite{zhang2015full,yang2025cooperative}.  
The RIS is composed of $N_{\text{s}}$ elements, each with a size of $\xi_{\text{s}} \times \xi_{\text{s}}$.  
Its phase shifts are denoted by $\boldsymbol{\omega} = [\omega_1, \omega_2, \ldots, \omega_{N_{\text{s}}}]^{\text{T}} \in \mathbb{C}^{N_{\text{s}} \times 1}$ and dynamically adjusted by the BS in several microseconds \cite{li2020intelligent}.
The positions of the TX, RX, and RIS are assumed to be known, whereas other environmental components are unknown but considered static.

The region of interest (ROI) is a predefined area where potential targets may be located.
It can be discretized into $N_{\text{v}}$ voxels, and the ROI image is represented as $\boldsymbol{\sigma} = [\sigma_1, \sigma_2, \ldots, \sigma_{N_{\text{v}}}]^{\text{T}} \in \mathbb{R}^{N_{\text{v}} \times 1}$, where $\sigma_{n_{\text{v}}}$ denotes the scattering coefficient of the $n_{\text{v}}$-th voxel, and $\sigma_{n_{\text{v}}} = 0$ if the voxel contains no objects.
The voxel side length is denoted by $\xi_{\text{v}}$, which is assumed to be much smaller than $\lambda/2$ \cite{mehrotra2022degrees,huang2024ris}.
It is assumed that $N_0$ symbol intervals are utilized to generate an image.
Although multiple subcarriers are available, the frequency index is omitted for simplicity, and the proposed methods can be readily extended to multi-subcarrier scenarios.

\subsection{Signal and Channel Models for Communication}

We assume that the communication user is equipped with a single antenna and is located either inside or outside the ROI.  
Denote $\mathbf{x} \in \mathbb{C}^{N_{\text{t}} \times 1}$ as the normalized transmitted signal at the TX and $P_{\text{t}}$ as the transmit power. The received signal at the user is given by  
\begin{equation}\label{eq-commun-signal}
{r}_{\text{com}} = \sqrt{P_{\text{t}}}\,\mathbf{h}_{\text{com-}\ast}^{\text{T}}\mathbf{x} + {n}_{\text{com}},
\end{equation} 
where $\mathbf{h}_{\text{com-}\ast}^{\text{T}}\in\mathbb{C}^{1\times N_{\text{t}}}$ represents the multipath channel from the TX to the user, and ${n}_{\text{com}}\in\mathbb{C}$ is the additive Gaussian noise.  
Here, $* \in \{\text{in}, \text{out}\}$ denotes the user position, corresponding to the channel for users inside or outside the ROI, respectively.  

According to Fig.~\ref{fig-model-in}, when the communication user is located inside the ROI, it simultaneously serves as the communication terminal and the target being sensed.
Thus, $\mathbf{h}_{\text{com-in}}$ is given by
\begin{equation}\label{eq-commun-channel-inside}
\begin{aligned}
\mathbf{h}_{\text{com-in}} = \mathbf{h}_{\text{tx-ue}} + \mathbf{h}_{\text{tx-ris-ue}} + \mathbf{h}_{\text{others}},
\end{aligned}
\end{equation}
where $\mathbf{h}_{\text{tx-ue}}=g_{\text{com}}\mathbf{h}'_{\text{tx-ue}}\in\mathbb{C}^{N_{\text{t}}\times 1}$ denotes the line-of-sight (LOS) path between the TX and the user.
Here, $g_{\text{com}}= {\lambda \sqrt{G_{\text{com}}}}/{\sqrt{4\pi}}$, and $G_{\text{com}}$ is the combined antenna gain of the TX and the user.  
The $n_{\text{t}}$-th element in $\mathbf{h}'_{\text{tx-ue}}$ is \cite{goldsmith2005wireless}
\begin{equation}\label{eq-tx-ue}
h'_{\text{tx-ue}, n_{\text{t}}} = \frac{e^{-j2\pi \frac{d_{n_{\text{t}}, {\text{ue}}}}{\lambda}}}{\sqrt{4\pi}d_{n_{\text{t}}, {\text{ue}}}},
\end{equation}
where $d_{n_{\text{t}}, {\text{ue}}}$ denotes the distance between the $n_{\text{t}}$-th TX antenna and the user.   
The RIS-reflected path is given by  
\begin{equation}\label{eq-commun-channel-single-bounce-ris}
\mathbf{h}_{\text{tx-ris-ue}} = g_{\text{com}}\mathbf{H}_{\text{tx-ris}} \text{diag}(\widetilde{\boldsymbol{\omega}}) \mathbf{h}_{\text{ris-ue}},
\end{equation}
where $\widetilde{\boldsymbol{\omega}} = g_{\text{ris}}\boldsymbol{\omega}$, and $g_{\text{ris}}=\sqrt{4\pi}\xi_{\text{s}}^2/\lambda$ is the scattering gain of the RIS elements \cite{huang2023joint}.
$\mathbf{H}_{\text{tx-ris}}^{\text{T}}$ and $\mathbf{h}_{\text{ris-ue}}^{\text{T}}$ denote the channels from the TX to the RIS and from the RIS to the user, respectively.  
The elements of both channels follow a similar form to \eqref{eq-tx-ue}.  
Finally, $\mathbf{h}_{\text{others}}$ includes multipath components involving more than two bounces and other scatterers. 

Alternatively, when the user is located outside the ROI, the target inside the ROI introduces additional multipaths into the communication channel, as depicted in Fig.~\ref{fig-model-out}.
In this case, $\mathbf{h}_{\text{com-out}}$ is given by   
\begin{multline} \label{eq-commun-channel-outside} 
\mathbf{h}_{\text{com-out}} =   \mathbf{h}_{\text{tx-ue}} + \mathbf{h}_{\text{tx-roi-ue}} + \mathbf{h}_{\text{tx-ris-ue}}\\
 + \mathbf{h}_{\text{tx-roi-ris-ue}} + \mathbf{h}_{\text{tx-ris-roi-ue}} + \mathbf{h}_{\text{others}}, 
\end{multline}
where the single-bounce path scattered by the ROI is given by   
\begin{equation}\label{eq-commun-channel-single-bounce}
\mathbf{h}_{\text{tx-roi-ue}} = g_{\text{com}}\mathbf{H}_{\text{tx-roi}} \text{diag}(\boldsymbol{\sigma}) \mathbf{h}_{\text{roi-ue}},
\end{equation}
with $\mathbf{H}_{\text{tx-roi}}^{\text{T}}$ and $\mathbf{h}_{\text{roi-ue}}^{\text{T}}$ denoting the channels from the TX to the ROI and from the ROI to the user, respectively.  
The twice-bounce paths are expressed as  
\begin{subequations}\label{eq-commun-channel-twice-bounce}
\begin{align}
&\mathbf{h}_{\text{tx-roi-ris-ue}} = g_{\text{com}}\mathbf{H}_{\text{tx-roi}} \text{diag}(\boldsymbol{\sigma}) \mathbf{H}_{\text{roi-ris}} \text{diag}(\widetilde{\boldsymbol{\omega}}) \mathbf{h}_{\text{ris-ue}}, \\
&\mathbf{h}_{\text{tx-ris-roi-ue}} = g_{\text{com}}\mathbf{H}_{\text{tx-ris}} \text{diag}(\widetilde{\boldsymbol{\omega}}) \mathbf{H}_{\text{ris-roi}} \text{diag}(\boldsymbol{\sigma}) \mathbf{h}_{\text{roi-ue}},
\end{align}
\end{subequations}
where $\mathbf{H}_{\text{roi-ris}}=\mathbf{H}_{\text{ris-roi}}^{\text{T}}$ is the channel between the ROI and the RIS.

\subsection{Signal and Channel Models for Sensing}
\label{sec-sensing-model}

The BS performs target sensing by transmitting signals from the TX and receiving the echoes at the RX.  
We assume that the RIS phases are altered using $K$ different configurations for imaging, given by $\boldsymbol{\Omega} = [\boldsymbol{\omega}_1, \boldsymbol{\omega}_2, \ldots, \boldsymbol{\omega}_K]^{\text{T}} \in \mathbb{C}^{K \times N_{\text{s}}}$.  
When employing the $k$-th RIS configuration $\boldsymbol{\omega}_k$ and transmitting the $n_{\text{t}}$-th signal $\mathbf{x}_{n_{\text{t}}}$, the received signal at the RX is expressed as 
\begin{equation}
\mathbf{r}_{\text{sen}} = \sqrt{P_{\text{t}}}\,{\mathbf{H}}_{\text{sen}, k}^{\text{T}}\mathbf{x}_{n_{\text{t}}} + \mathbf{n}_{\text{sen}},
\end{equation}
where $\mathbf{H}_{\text{sen}, k}\in\mathbb{C}^{N_{\text{t}}\times N_{\text{r}}}$ denotes the multipath channel between the TX and the RX, and $\mathbf{n}_{\text{sen}} \in \mathbb{C}^{N_{\text{r}}}$ represents the additive noise at the RX.   

According to Fig.~\ref{fig-model}, ${\mathbf{H}}_{\text{sen}, k}$ can be expressed as
\begin{multline} \label{eq-sensing-channel}
{\mathbf{H}}_{\text{sen}, k} =  \mathbf{H}_{\text{tx-roi-rx}} + \mathbf{H}_{\text{tx-ris-rx}, k} \\
+ \mathbf{H}_{\text{tx-roi-ris-rx}, k} + \mathbf{H}_{\text{tx-ris-roi-rx}, k} + {\mathbf{H}}_{\text{others}, k}, 
\end{multline}
\hspace{-0.5em}where $\mathbf{H}_{\text{tx-roi-rx}}$ and $\mathbf{H}_{\text{tx-ris-rx}, k}$ are the single-bounce paths scattered by the ROI and RIS, respectively, given by 
\begin{subequations}\label{eq-sen-channel-single-bounce}
\begin{align}
\mathbf{H}_{\text{tx-roi-rx}} & = g_{\text{sen}}\mathbf{H}_{\text{tx-roi}} \text{diag}(\boldsymbol{\sigma}) \mathbf{H}_{\text{roi-rx}}, \\
\mathbf{H}_{\text{tx-ris-rx},k} & = g_{\text{sen}}\mathbf{H}_{\text{tx-ris}} \text{diag}(\widetilde{\boldsymbol{\omega}}_k) \mathbf{H}_{\text{ris-rx}},
\end{align}
\end{subequations}
where $g_{\text{sen}}={\lambda \sqrt{G_{\text{sen}}}}/{\sqrt{4\pi}}$, and $G_{\text{sen}}$ is the combined antenna gain of the TX and the RX.
$\mathbf{H}_{\text{tx-roi-ris-rx}, k}$ and $\mathbf{H}_{\text{tx-ris-roi-rx}, k}$ represent the twice-bounce paths, given by  
\begin{subequations}\label{eq-sen-channel-twice-bounce}
\begin{align} 
\mathbf{H}_{\text{tx-roi-ris-rx},k} &= g_{\text{sen}}\mathbf{H}_{\text{tx-roi}} \text{diag}(\boldsymbol{\sigma}) \mathbf{H}_{\text{roi-ris}} \text{diag}(\widetilde{\boldsymbol{\omega}}_k) \mathbf{H}_{\text{ris-rx}}, \\
\mathbf{H}_{\text{tx-ris-roi-rx},k} &= g_{\text{sen}}\mathbf{H}_{\text{tx-ris}} \text{diag}(\widetilde{\boldsymbol{\omega}}_k) \mathbf{H}_{\text{ris-roi}} \text{diag}(\boldsymbol{\sigma}) \mathbf{H}_{\text{roi-rx}}.
\end{align}
\end{subequations} 
The term ${\mathbf{H}}_{\text{others}, k}$ includes multipath components involving more than two bounces or scatterers beyond the ROI and the RIS, representing the unknown portion of ${\mathbf{H}}_{\text{sen}, k}$.

The ROI image $\boldsymbol{\sigma}$ is expected to be inferred from ${\mathbf{H}}_{\text{sen}, k}$, since this channel includes components scattered by the ROI and thus encodes the image information.  
Moreover, as RIS phases influence the paths scattered by both the ROI and the RIS, altering the RIS configurations provides diverse sensing perspectives toward the target, significantly increasing the effective number of measurements.

\subsection{ISAC Protocol}

\begin{figure}[t]
    \centering
    \includegraphics[width=0.9\linewidth]{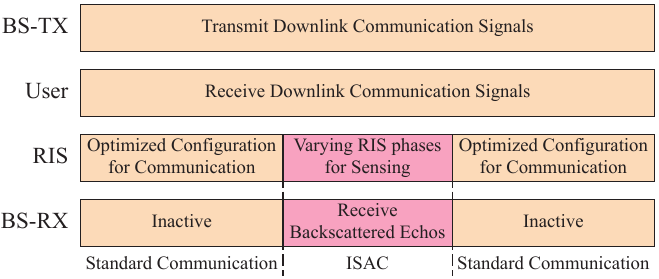}
    \captionsetup{font=footnotesize}
    \caption{Illustration of the proposed ISAC protocol.}
    \label{fig-protocol}
\end{figure}

The protocol of the ISAC system is depicted in Fig. \ref{fig-protocol}.
The TX and the user keep uninterrupted communication by transmitting and receiving downlink communication signals, respectively.
During standard communication phases, RIS phases are optimized to enhance communication performance, whereas the RX does not work.
Alternatively, during the ISAC interval, the RIS phases are varied across $K$ unique configurations, while the RX receives backscattered echoes, achieving \textit{communication-assisted sensing} by multiplexing communication signals.
While RIS phase optimization for imaging is explored in \cite{huang2024ris,hu2022metasketch}, this study concentrates on the imaging algorithm design, assuming that $\boldsymbol{\Omega}$ is randomly generated beforehand.
Subsequently, the formulated ROI images can be used to refine communication-oriented RIS phase designs, thereby enhancing subsequent communication performance and realizing \textit{sensing-assisted communication}, as elaborated in Sec. \ref{sec-imaging-augmented-communication}.

\section{INR-Based Wireless Imaging}
\label{sec-inr-imaging}

INR is a novel technique grounded in physical modeling and implemented using DL techniques.
This section first presents the essential CSI measurements and forward model used by INR.
Then, the network structure and INR-based imaging algorithm design are presented subsequently.
Finally, the advantages of INR are highlighted by comparing it with traditional model-based and data-driven DL methods.

\subsection{CSI Measurements and Forward Model}

INR is developed under the supervision of CSI measurements and forward models, as described in \eqref{eq-general-model}.  
This subsection explains how they are obtained in the considered ISAC system. 

\subsubsection{CSI Measurements}
In this study, ${\mathbf{H}}_{\text{sen}, k}$ is first estimated before performing imaging.  
Under the least squares (LS) channel estimation algorithm, $N_{\text{t}}$ signals are required \cite{tang2023joint}.  
\footnote{
The number of measurements required to estimate ${\mathbf{H}}_{\text{sen}, k}$ can be reduced well below $N_{\text{t}}$ by leveraging the sparse nature of ${\mathbf{H}}_{\text{sen}, k}$ \cite{tang2023joint}. In this study, we adopt the LS method as a simple example.}  
The transmit signals are denoted as $\mathbf{X} = [\mathbf{x}_{1}, \mathbf{x}_{2}, \ldots, \mathbf{x}_{N_{\text{t}}}]$, which are known at the BS and can be designed using dedicated precoding techniques \cite{zheng2024random}.  
The estimated CSI is then expressed as
\begin{equation}\label{eq-sensing-measurement}
\widehat{\mathbf{H}}_{\text{sen}, k} = {\mathbf{H}}_{\text{sen}, k} + \mathbf{N}_{\text{sen}, k},
\end{equation}
where $\mathbf{N}_{\text{sen}, k}$ denotes additional noise caused by channel estimation errors.  
As a result, $N_0 = K N_{\text{t}}$ symbol intervals are used to acquire sufficient CSI measurements $\{\widehat{\mathbf{H}}_{\text{sen}, k}\}_{k=1}^{K}$.  
Based on the general model in \eqref{eq-general-model}, the measurement vector $\mathbf{y}$ is formed by stacking all elements in $\{\widehat{\mathbf{H}}_{\text{sen}, k}\}_{k=1}^{K}$ into a single vector of dimension $N_{\text{m}} = K N_{\text{t}} N_{\text{r}}$.
In contrast to prior studies \cite{huang2024fourier,huang2024ris,hu2022metasketch}, this approach eliminates the need for multipath extraction and simplifies system design.

\subsubsection{Forward Model}
The physical relationship between CSI measurements and the ROI image must be leveraged to impose constraints during image reconstruction.  
We assume that all point-to-point channels remain constant under the quasi-static condition.   
In \eqref{eq-sensing-channel}, the term ${\mathbf{H}}_{\text{others}, k}$ represents unknown channel components due to limited knowledge of the environment, whereas other components are explicitly modeled in \eqref{eq-sen-channel-single-bounce} and \eqref{eq-sen-channel-twice-bounce}.  
Thus, ${\mathbf{H}}_{\text{others}, k}$ is treated as disturbance to a partially known physical model \cite{huang2024ris,huang2024fourier,hu2022metasketch}.  
Accordingly, \eqref{eq-sensing-measurement} can be reformulated as  
\begin{equation}\label{eq-sensing-measurement-new0}
\widehat{\mathbf{H}}_{\text{sen}, k} = {\mathbf{H}}'_{\text{sen}, k} + \mathbf{N}'_{\text{sen}, k},
\end{equation}
where 
\begin{equation}\label{eq-sensing-measurement-new}
\begin{aligned}
& {\mathbf{H}}'_{\text{sen}, k} = \mathbf{H}_{\text{tx-roi-rx}} + \mathbf{H}_{\text{tx-ris-rx}, k} + \mathbf{H}_{\text{tx-roi-ris-rx}, k} + \mathbf{H}_{\text{tx-ris-roi-rx}, k},\\
& \mathbf{N}'_{\text{sen}, k} = {\mathbf{H}}_{\text{others}, k} + \mathbf{N}_{\text{sen}, k}.
\end{aligned}
\end{equation}
Consequently, the forward model corresponding to the $k$-th RIS configuration $\boldsymbol{\omega}_k$ is expressed as $f_k(\boldsymbol{\sigma}, \boldsymbol{\omega}_k) = \text{vec}({\mathbf{H}}'_{\text{sen}, k})$, where $\text{vec}(\cdot)$ denotes the vectorization of a matrix.  
By stacking all $K$ components in $\{f_k(\boldsymbol{\sigma}, \boldsymbol{\omega}_k)\}_{k=1}^{K}$, the complete forward model is denoted as $f(\boldsymbol{\sigma}, \boldsymbol{\Omega})$.
Note that the forward physical model can be modified with available environmental information, e.g., occlusions and reflections.

\begin{remark}
The proposed INR-based wireless imaging method is potentially generalizable to diverse imaging systems, provided that their corresponding CSI measurements and forward physical models are available.
The only requirement for physical models is differentiable.
To illustrate this applicability, this study investigates the wireless imaging problem within the context of RIS-aided ISAC systems.
\end{remark}

\subsection{Network Structure}
\label{subsec-network-structure}

The INR technique employs an NN to learn the representation of the continuous shape and scattering characteristics of the ROI image.  
Following advanced DL methods \cite{mildenhall2021nerf}, the input to the INR is any position $\mathbf{p} = [x, y, z]^{\text{T}}$ within the ROI, and its output is the corresponding image parameter at that location, i.e., the scattering coefficient at $\mathbf{p}$, denoted as $\sigma_{\mathbf{p}} \in \mathbb{R}$.  
Let the INR be denoted by $\mathcal{M}_{\boldsymbol{\theta}}$ with trainable parameters $\boldsymbol{\theta}$; then, we have  
\begin{equation}
\mathcal{M}_{\boldsymbol{\theta}}\ :\ \mathbf{p} \longrightarrow \sigma_{\mathbf{p}}.
\end{equation}
Thus, the continuous ROI image is implicitly embedded into $\mathcal{M}_{\boldsymbol{\theta}}$ through its specific NN architecture and parameters.

The network structure of the INR is typically a multi-layer perceptron (MLP).  
In this study, we adopt an MLP consisting of six fully connected (FC) layers, each containing 256 neurons.  
However, MLPs tend to exhibit a bias toward learning low-frequency functions, making it difficult to capture high-frequency variations \cite{chatelier2025model}.  
Note that the term \textit{high-/low-frequency} here refers to the variation rate of voxel values with respect to their locations, rather than the frequency of wireless signals.  
The ROI image typically includes a large number of high-frequency components, especially near targets' edges, where small movements in position $\mathbf{p}$ can lead to significant changes in voxel values.  
To address this limitation, specific positional encoding and a designated activation function are introduced to enhance the MLP’s capability in learning the continuous ROI image \cite{chatelier2025model}.

\begin{figure*}[t]
    \centering
    \includegraphics[width=0.95\linewidth]{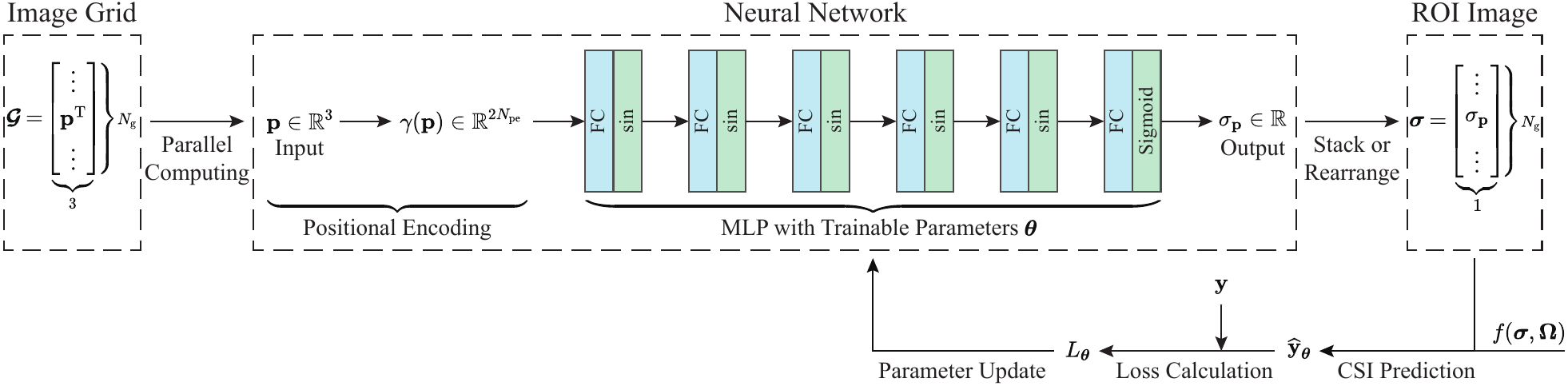}
    \captionsetup{font=footnotesize}
    \caption{Illustration of the INR network structure and the online training process.}
    \label{fig-network}
\end{figure*}

\subsubsection{Positional Encoding}

Although the INR maps the 3D position $\mathbf{p}$ to its corresponding image parameter, $\mathbf{p}$ is not a suitable direct input to the MLP for two reasons.  
First, MLPs are generally more effective at learning mappings from high-dimensional inputs to low-dimensional outputs, while $\mathbf{p}$ is low-dimensional and thus contains limited information.  
Second, position variation in the image is inherently low-frequency, meaning $\mathbf{p}$ itself lacks the high-frequency components beneficial for INR-based imaging.  
Therefore, $\mathbf{p}$ should be transformed into a higher-dimensional representation that captures more frequency components. This process is referred to as \textit{positional encoding}. 

Although various positional encoding techniques have been utilized in literature \cite{molaei2023implicit}, this work does not aim to provide a comprehensive review.  
Instead, we adopt an effective method tailored to the wireless imaging scenario.  
Specifically, we employ Fourier feature mapping \cite{tancik2020fourier}, defined as   
\begin{equation}
\boldsymbol{\gamma}(\mathbf{p}) = \left[\begin{array}{c}\cos(2\pi \mathbf{Bp}) \\ \sin(2\pi \mathbf{Bp})\end{array}\right],
\end{equation}
where Fourier features are represented by $\sin$ and $\cos$ functions. The matrix 
$\mathbf{B}\in\mathbb{R}^{N_{\text{pe}}\times 3}$ is a random Gaussian matrix, with elements independently sampled from a Gaussian distribution $\mathcal{N}(0, \chi^2)$.
The randomness in $\mathbf{B}$ enables $\boldsymbol{\gamma}(\mathbf{p})$ to include components across a broad frequency spectrum.  
Here, $N_{\text{pe}}$ is the dimensionality of positional encoding, and $\chi$ is a hyperparameter controlling the variance of the elements in $\mathbf{B}$.

\subsubsection{Activation Function}

The rectified linear unit (ReLU) is commonly used as the activation function for FC layers.  
However, due to its piecewise linear nature and lack of second-order derivatives, ReLU performs poorly in learning high-frequency signals.  
For instance, it is not well suited to approximating a sine function.  
Alternatively, continuous and periodic functions have been shown to be effective as activation functions in INR-based architectures \cite{sitzmann2020implicit}.  
In this study, we replace ReLU with the $\sin$ function to improve the MLP’s ability to learn high-frequency content.  
The $i$-th hidden layer can be expressed as   
\begin{equation}
\phi_i(\mathbf{x}_{i}) = \sin\left(\kappa\left(\mathbf{W}_i\mathbf{x}_i+\mathbf{b}_i\right)\right),
\end{equation}
where $\mathbf{x}_i$, $\mathbf{W}_i$, and $\mathbf{b}_i$ represent the input, weights, and biases of the $i$-th layer, respectively.  
The parameter $\kappa$ is a hyperparameter controlling the periodicity of the sine activation.  
As a result, the continuous image features and high-frequency components are better preserved and recovered within the MLP.  
However, sine-based non-linearities can lead to high sensitivity to the NN's parameter initialization.  
Therefore, we apply the initialization strategy proposed in \cite{sitzmann2020implicit}.  
Finally, the output layer employs the Sigmoid activation function, and the overall INR network architecture is illustrated in Fig.~\ref{fig-network}.

\subsection{NN Training and Image Generation}
\label{subsec-nn-training}

INR extracts image information from CSI measurements with the aid of the physical model through NN training, where this information is implicitly embedded into NN parameters, rather than directly or explicitly presented as an image with discrete voxels.
To achieve this, INR transforms the direct image-domain optimization in Problem (P1) into an indirect parameter-domain optimization, given by
\begin{equation*}
\text{(P2)}\quad  \boldsymbol{\theta}^\star = \mathop{\arg\!\min}\limits_{\boldsymbol{\theta}} \mu(f(\mathcal{M}_{\boldsymbol{\theta}}(\boldsymbol{\mathcal{G}}^{\circ}), \boldsymbol{\Omega}), \mathbf{y}) + \rho(\mathcal{M}_{\boldsymbol{\theta}}(\boldsymbol{\mathcal{G}}^{\circ})),\\
\end{equation*}
where $\boldsymbol{\mathcal{G}}^{\circ}\in\mathbb{R}^{N_{\text{v}}\times 3}$ denotes the discrete sampling grid of the ROI and determines the resolution of the final predicted image.  

Based on (P2), the NN is trained to represent the ROI image under the constraints imposed by the CSI measurements $\mathbf{y}$ and the forward model $f(\boldsymbol{\sigma}, \boldsymbol{\Omega})$.
In addition to the explicit regularization term $\rho(\mathcal{M}_{\boldsymbol{\theta}}(\boldsymbol{\mathcal{G}}^{\circ}))$, Problem (P2) also leverages an implicit image-statistics prior encoded by the NN structure $\mathcal{M}_{\boldsymbol{\theta}}$, meaning that all position-parameter pairs $(\mathbf{p}, \sigma_{\mathbf{p}})$ share a common continuous underlying function \cite{shen2022nerp,mildenhall2021nerf}.
Diverse from traditional data-driven DL methods, the NN training and inference procedures of the proposed INR-based imaging algorithm are both executed in an online manner.
Specifically, the online training procedure iteratively follows four steps:

\subsubsection{Step 1---NN Inference}\label{subsub-step-1-nn-inference}
In each epoch, INR transforms the image information implicitly embedded in the NN parameters to an explicit image by performing inference on the training grid $\boldsymbol{\mathcal{G}} \in \mathbb{R}^{N_{\text{g}} \times 3}$, which contains $N_{\text{g}}$ point locations within the ROI.\footnote{
In practical implementation, we can set $\boldsymbol{\mathcal{G}}=\boldsymbol{\mathcal{G}}^\circ$ and $N_{\text{g}}=N_{\text{v}}$ to use the same image resolution for NN training and final image prediction.
This uniform grid can be calculated based on the known ROI boundaries and selected sampling interval \cite{hu2022metasketch,tong2025computational,huang2024ris}.
However, $\boldsymbol{\mathcal{G}}$ can be different from $\boldsymbol{\mathcal{G}}^\circ$, since INR employs a continuous function, i.e., the MLP, to represent the ROI image, thus allowing arbitrary sampling grid density.
A dense grid $\boldsymbol{\mathcal{G}}$ enables the NN to learned fine-grained image details, whereas a sparse grid can accelerate NN training.}
Moreover, $\boldsymbol{\mathcal{G}}$ can be non-uniformly sampled to form denser grids in regions of targets or sparser grids in empty areas \cite{mildenhall2021nerf}.
As illustrated by the dashed boxes in Fig. \ref{fig-network}, the NN receives the grid locations $\boldsymbol{\mathcal{G}}$ as the input and outputs the corresponding predicted scattering coefficients:
\begin{equation}\label{eq-training-1}
\widehat{\boldsymbol{\sigma}}_{\boldsymbol{\theta}} = \mathcal{M}_{\boldsymbol{\theta}}(\boldsymbol{\mathcal{G}}).
\end{equation}
By rearranging the elements in $\widehat{\boldsymbol{\sigma}}_{\boldsymbol{\theta}}$ based on their spatial positions defined by $\boldsymbol{\mathcal{G}}$, an ROI image is generated, whose resolution equals the sampling interval of $\boldsymbol{\mathcal{G}}$.
Note that the NN can generate an image even without training, since its input $\boldsymbol{\mathcal{G}}$ has been determined.
However, the derived image may be significantly distorted and represents little GT image information, due to under-optimized NN parameters.
During training, the NN repetitively generates predicted images in each epoch, which is indirectly supervised by CSI measurements in subsequent steps.

\subsubsection{Step 2---CSI Prediction}
Given the predicted image $\widehat{\boldsymbol{\sigma}}_{\boldsymbol{\theta}}$, the corresponding CSI is computed using the forward physical model:   
\begin{equation}\label{eq-training-2}
\widehat{\mathbf{y}}_{\boldsymbol{\theta}} = f(\widehat{\boldsymbol{\sigma}}_{\boldsymbol{\theta}}, \boldsymbol{\Omega}).
\end{equation}
$\widehat{\mathbf{y}}_{\boldsymbol{\theta}}$ is the CSI if $\widehat{\boldsymbol{\sigma}}_{\boldsymbol{\theta}}$ was the GT image.
This embeds physics domain knowledge into the learning procedure \cite{guo2023physics,yu2024ai}.

\subsubsection{Step 3---Loss Calculation}
The NN is trained to minimize the discrepancy between the CSI $\widehat{\mathbf{y}}_{\boldsymbol{\theta}}$ corresponding to the predicted image $\widehat{\boldsymbol{\sigma}}_{\boldsymbol{\theta}}$ and the CSI $\mathbf{y}$ measured by the ISAC system.  
The loss function is defined as  
\begin{equation}\label{eq-training-3}
L_{\boldsymbol{\theta}} = \|\widehat{\mathbf{y}}_{\boldsymbol{\theta}} - {\mathbf{y}}\|_2 + \alpha \|\widehat{\boldsymbol{\sigma}}_{\boldsymbol{\theta}}\|_1,
\end{equation}
where the first term enforces CSI consistency (the data term in (P2)), and the second term is an explicit regularizer promoting sparsity in the ROI image.  
The hyperparameter $\alpha$ controls the weight of the sparsity constraint.
Here, the NN output in \eqref{eq-training-1} is not directly supervised, since its GT value is unknown; instead, it is transformed from the image domain to the CSI domain in \eqref{eq-training-2} and indirectly supervised by CSI measurements in \eqref{eq-training-3}.

\begin{table*}[t]
    \renewcommand{\arraystretch}{1.3}
    \renewcommand\cellgape{\Gape[2pt]}
    \centering
    \fontsize{8}{8}\selectfont
    \captionsetup{font=small}
    \caption{Comparison between model-based FT, CS, data-driven DL, and INR imaging techniques.}\label{tab-compare}
    \begin{threeparttable}
    \begin{tabular}{|c|P{3cm}<{\centering}|P{3cm}<{\centering}|P{3cm}<{\centering}|P{3.2cm}<{\centering}|}
            \hline
             & FT \cite{sheen2001three,huang2024fourier} & CS \cite{hu2022metasketch,huang2024ris,tong2025computational,lyu2024compressed,huang2025cooperative} & Data-driven DL \cite{lu2024deep,li2020intelligent,qi2023resource} & INR \\
            \hline
            Measurement amount & \makecell{Large, satisfying \\ Nyquist criterion} & Small, using sparse prior & \makecell{Small, using \\ training data prior} & \makecell{Small, using sparse prior \\ and DL-based optimization} \\
            \hline
            Multipath extraction & Request & Request & None & None \\
            \hline
            Forward model requirement & Spatial integral & Sensing matrix & None & Any differentiable model \\
            \hline
            Image resolution dependence & FFT points & \makecell{Predefined voxel gird \\ in the sensing matrix} & \makecell{NN structure and \\ training data} & Arbitrary \\
            \hline
            ``CSI-image'' dataset amount & None & None & Extremely Large & None \\
            \hline
            Imaging target requirement & None & Sparse & Same distribution with training data & None \\
            \hline
            Computational platform & CPU & CPU & GPU & GPU \\
            \hline
            Imaging procedure & \makecell{FFT and \\ Hadamard product} & Iterative CS algorithm & Offline training and online inference & \makecell{Online training and \\ image generation} \\
            \hline
        \end{tabular}
    \end{threeparttable}
\end{table*}

\subsubsection{Step 4---NN Parameter Update}
The NN parameters are updated according to the gradient of the loss function:
\begin{equation}
\boldsymbol{\theta} \longleftarrow \boldsymbol{\theta}-r\nabla_{\boldsymbol{\theta}} L_{\boldsymbol{\theta}}  ,
\end{equation}
where $r$ is the learning rate and $\nabla_{\boldsymbol{\theta}} L_{\boldsymbol{\theta}}$ denotes the gradient of $L_{\boldsymbol{\theta}}$ with respect to $\boldsymbol{\theta}$.  
This update can be implemented using standard DL frameworks such as PyTorch or TensorFlow.

The above online training steps have been depicted in Fig. \ref{fig-network}, requesting only one piece of CSI measurements $\mathbf{y}$.
Moreover, Fig. \ref{fig-network} has been simplified into the dashed box of Fig. \ref{fig-compare}(c), together with the following online image generation procedure.

After the NN is trained, its parameters have been optimized to implicitly represent the ROI image.
To transform this implicitly embedded information into explicit images, we can reapply Step 1 in Sec. \ref{subsub-step-1-nn-inference} using the desired image grid $\boldsymbol{\mathcal{G}}^{\circ}$ as input.
This image generation procedure is given as\footnote{
The CSI measurements $\mathbf{y}$ and the forward physical model $f(\boldsymbol{\sigma}, \boldsymbol{\Omega})$ are used to assist in NN training.
After the training procedure, their associated image information has been extracted and embedded into NN parameters.
Thus, the final image generation step does not utilize $\mathbf{y}$ and $f(\boldsymbol{\sigma}, \boldsymbol{\Omega})$.}
\begin{equation} 
\widehat{\boldsymbol{\sigma}} =\mathcal{M}_{\boldsymbol{\theta}^\star}(\boldsymbol{\mathcal{G}}^{\circ}),
\end{equation}
where $\boldsymbol{\theta}^\star$ denote optimized NN parameters.
Although the training grid $\boldsymbol{\mathcal{G}}$ is discretized, the NN implicitly learns and encodes a continuous representation of the ROI image.  
Therefore, INR allows flexible output image resolutions by adjusting the sampling density of $\boldsymbol{\mathcal{G}}^{\circ}$ during final image generation.
Thus, voxel values off the training grid $\boldsymbol{\mathcal{G}}$ can be obtained, which is infeasible for traditional model-based or data-driven imaging approaches.
This ability is presented by the simulation results in Sec. \ref{subsec-result-threeD}.

\subsection{Comparison with Traditional Imaging Algorithms}
\label{subsec-compare}

\begin{figure}[t]
    \centering
    \includegraphics[width=\linewidth]{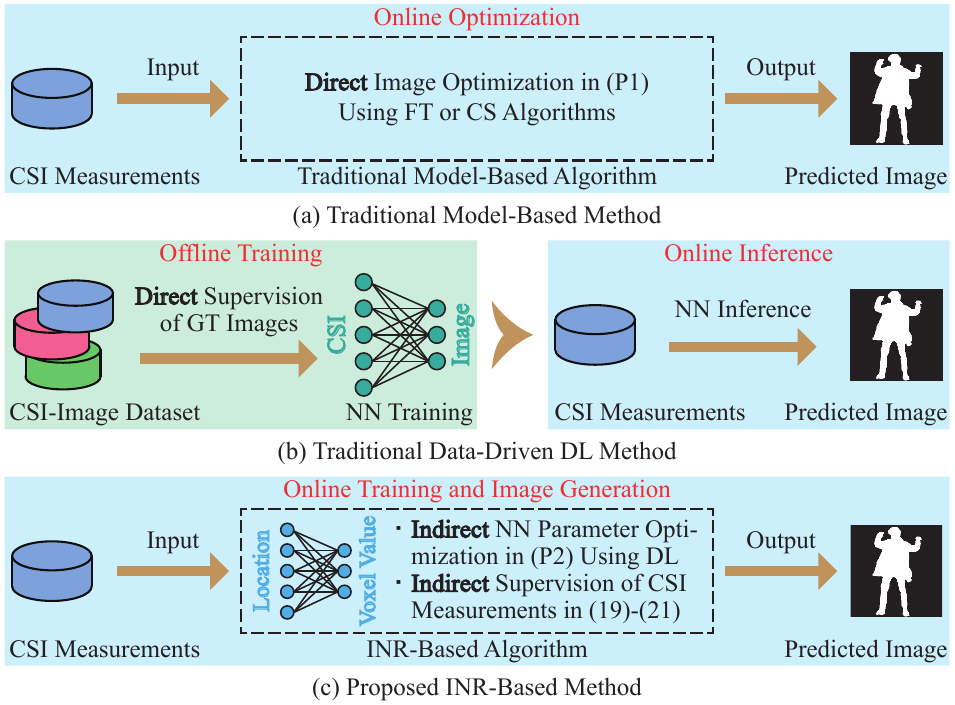}
    \captionsetup{font=footnotesize}
    \caption{Imaging mechanisms of model-based, data-driven DL, and INR-based methods.}
    \label{fig-compare}
\end{figure}

This subsection compares INR with traditional model-based algorithms and data-driven DL techniques, as illustrated in Fig. \ref{fig-compare}.
Traditional data-driven DL methods require an offline training stage to learn the projection function from CSI measurements to images, demanding a large ``CSI-image'' dataset \cite{lu2024deep,li2020intelligent,qi2023resource}.
Alternatively, traditional model-based algorithms and the proposed INR-based method only involve online operations, extracting the image information based on the constraints posed by the physical model, rather than from the statistical features of numerous ``CSI-image'' pairs.
Instead of being directly supervised by GT images, INR-based imager transforms the NN outputs to the CSI domain, which is indirectly supervised by CSI measurements, as detailed by \eqref{eq-training-1}, \eqref{eq-training-2}, and \eqref{eq-training-3}.
Thus, the INR-based method does not require constructing large ``CSI-image'' training datasets nor requesting that test target images possess the same statistical distribution with training data.
This makes INR possess high data efficiency and generalization.

The data requirement of INR is the same as traditional model-based algorithms.
However, the direct model-based image optimization Problem (P1) has been transformed to indirect NN parameter optimization Problem (P2) in INR, since the target image is implicitly represented by the NN and cannot be directly optimized.
Nevertheless, the continuous image representation using the NN enables INR to generate arbitrary-resolution images.
Additionally, DL-based optimization in INR allows it to support arbitrary differentiable forward models, whereas FT and CS approaches depend on predefined, specific model formulations \cite{hu2022metasketch,huang2024ris,tong2025computational,lyu2024compressed,huang2025cooperative,sheen2001three,huang2024fourier}.
The above features make the proposed INR-based imager significantly superior to previous techniques, and the detailed comparison is enumerated in Table \ref{tab-compare}.

Nevertheless, additional issues of INR should be addressed.  
First, the imaging performance of INR strongly depends on the accuracy of the forward physical model, which guides NN parameter optimization.  
This issue is also critical for traditional model-based imaging methods \cite{huang2024ris,hu2022metasketch,tong2025computational}.  
However, INR partially mitigates this problem by leveraging the powerful learning capabilities of DL, as discussed in Sec.~\ref{subsec-enhanced-imager-multipath}.  
Second, INR requires online NN training for each imaging target, which introduces a delay in producing the results.  
By contrast, traditional DL methods allow near real-time inference after offline training.  
However, our simulations show that INR typically completes training within several seconds, which is considerably faster than the hours-long training time required by conventional DL models \cite{lu2024deep,li2020intelligent,qi2023resource}.  
Furthermore, the use of prior information can accelerate the NN training process, as introduced in Sec.~\ref{subsec-enhanced-imager-dynamic}.

\section{Enhanced Imager in Practical Environments}
\label{sec-enhanced-imager}

This section addresses the challenges of multipath interference and dynamic target imaging to enhance the performance of INR in practical environments. 

\subsection{Imaging under Multipath Interference}
\label{subsec-enhanced-imager-multipath}

In Sec.~\ref{sec-inr-imaging}, the term ${\mathbf{H}}_{\text{others}, k}$ in \eqref{eq-sensing-channel} is considered as a disturbance to the forward physical model, as shown in \eqref{eq-sensing-measurement-new}, due to the unknown characteristics of surrounding environments.  
However, the performance of INR-based imaging deteriorates when the energy of ${\mathbf{H}}_{\text{others}, k}$ becomes large.  
To enable robust ROI imaging under severe multipath interference, we propose a two-step strategy that involves background calibration and residual interference learning.

\subsubsection{Background Calibration}
\label{subsubsec-channel-calibration}

Revisiting the sensing channel model in \eqref{eq-sensing-channel}, ${\mathbf{H}}_{\text{others}, k}$ includes multipath components scattered by background elements, such as TX-background-RX, TX-background-RIS-RX, and TX-background-ROI-RX.  
Therefore, the multipaths in ${\mathbf{H}}_{\text{sen}, k}$ can be divided into ROI-related and ROI-irrelated components.  
The latter includes paths scattered only by the RIS and background scatterers.  
To suppress their interference, ROI-irrelated components can be measured when no targets are present in the ROI and subtracted from the CSI measurements $\widehat{\mathbf{H}}_{\text{sen}, k}$ during actual imaging process.
This approach, known as background calibration \cite{li2024radio}, results in a residual channel given by  
\begin{equation}\label{eq-joint-learning-channel}
{\mathbf{H}}_{\text{ima}, k} = \underbrace{\mathbf{H}_{\text{roi-bg}}}_{\text{unknown}} + \underbrace{\mathbf{H}_{\text{tx-roi-rx}} + \mathbf{H}_{\text{tx-roi-ris-rx}, k} + \mathbf{H}_{\text{tx-ris-roi-rx}, k}}_{\text{can be modeled}},
\end{equation}
where $\mathbf{H}_{\text{roi-bg}}$ denotes multipath scattered by both the ROI and background scatterers.  
The calibrated CSI measurement is
\begin{equation}\label{eq-calibration-channel-measurement}
\widehat{\mathbf{H}}_{\text{ima}, k} = {\mathbf{H}}_{\text{ima}, k} + \mathbf{N}_{\text{ima}},
\end{equation}
where $\mathbf{N}_{\text{ima}}$ includes both channel estimation and calibration noise.
In practice, calibration noise can be significantly reduced by averaging multiple measurements taken in the absence of targets.  
The final measurement vector for imaging is then reformulated as $\mathbf{y} = \text{vec}(\{\widehat{\mathbf{H}}_{\text{ima}, k}\}_{k=1}^K)$.

\subsubsection{Joint Imaging and Multipath Interference Learning}
\label{subsubsec-joint-multipath-learning}

Although most unknown multipath components are removed via calibration, the forward physical model remains partially known, as described in \eqref{eq-joint-learning-channel}.  
Following prior work \cite{hu2022metasketch,huang2024ris,tong2025computational}, the unknown component $\mathbf{H}_{\text{roi-bg}}$ can be approximated as
\begin{equation}
\text{vec}({\mathbf{H}_{\text{roi-bg}}}) = \mathbf{A}\boldsymbol{\sigma},
\end{equation}
where $\mathbf{A}\in\mathbb{C}^{N_{\text{t}}N_{\text{r}}\times N_{\text{v}}}$ is an unknown matrix equivalently representing environmental multipath propagations.
Therefore, the partially known forward model for the $k$-th RIS configuration becomes
\begin{multline}  
f'_k(\boldsymbol{\sigma}, \boldsymbol{\omega}_k, \mathbf{A}) = \mathbf{A}\boldsymbol{\sigma} + \text{vec}(\mathbf{H}_{\text{tx-roi-rx}} +  \mathbf{H}_{\text{tx-roi-ris-rx}, k} \\ + 
 \mathbf{H}_{\text{tx-ris-roi-rx}, k}). 
\end{multline}
By stacking $\{f'_k(\boldsymbol{\sigma}, \boldsymbol{\omega}_k, \mathbf{A})\}_{k=1}^K$, we obtain the complete model $f'(\boldsymbol{\sigma}, \boldsymbol{\Omega}, \mathbf{A})$.
To mitigate the influence of the unknown term $\mathbf{H}_{\text{roi-bg}}$, the matrix $\mathbf{A}$ is treated as a learnable parameter, and (P2) is reformulated as 
\begin{multline*} 
\text{(P3)}\ (\boldsymbol{\theta}^\star, \mathbf{A}^\star) = \mathop{\arg\!\min}\limits_{\boldsymbol{\theta}, \, \mathbf{A}} \ \mu(f'(\mathcal{M}_{\boldsymbol{\theta}}(\boldsymbol{\mathcal{G}}^{\circ}), \boldsymbol{\Omega}, \mathbf{A}), \mathbf{y}) \\ 
+ \rho(\mathcal{M}_{\boldsymbol{\theta}}(\boldsymbol{\mathcal{G}}^{\circ})), 
\end{multline*}
which can be solved using the training framework described in Sec.~\ref{subsec-nn-training}.  
Under the supervision of this augmented forward model, both the image representation $\mathcal{M}_{\boldsymbol{\theta}}$ and the interference term $\mathbf{A}$ are jointly optimized, thus enabling robust imaging in the presence of multipath interference.

\subsection{Successive Imaging for Dynamic Targets}
\label{subsec-enhanced-imager-dynamic}

INR is considered an effective imaging technique due to the advantages discussed in Sec.~\ref{subsec-compare}.  
However, the latency caused by online NN training poses a challenge for dynamic target sensing, where a high image update rate is often required.  
Although prior work has proposed sophisticated strategies to accelerate training \cite{mildenhall2021nerf,muller2022instant}, we draw inspiration from medical imaging methods \cite{shen2022nerp} and propose a simple yet effective solution by incorporating prior image information into the training process.

Specifically, we assume that the shape and position of the targets within the ROI are static during one imaging procedure (at a single time instant) but evolve smoothly over time instants \cite{yang2025cooperative}.
This implies a strong temporal correlation between ROI images at adjacent time steps, allowing historical images to serve as valuable priors.  
Suppose the image at time $t-1$ is represented by $\mathcal{M}_{\boldsymbol{\theta}_{t-1}}$.  
Then, instead of using only a regularization term $\rho(\cdot)$ as in (P2), we embed prior information from $\mathcal{M}_{\boldsymbol{\theta}_{t-1}}$ into the imaging formulation at time $t$, yielding   
\begin{equation*}
\text{(P4)}\quad  \boldsymbol{\theta}_t^\star = \mathop{\arg\!\min}\limits_{\boldsymbol{\theta}_t} \mu(f(\mathcal{M}_{\boldsymbol{\theta}_t}(\boldsymbol{\mathcal{G}}^{\circ}), \boldsymbol{\Omega}), \mathbf{y}; \mathcal{M}_{\boldsymbol{\theta}_{t-1}}),
\end{equation*}
where the prior image is used to initialize the parameters of the new network.  
The optimization is then performed following the same procedure as described in Sec.~\ref{subsec-nn-training}.  
This prior-based initialization yields two significant benefits.  
First, imaging speed is improved, since optimization starts from an already close-to-optimal status.  
Second, imaging quality is enhanced through knowledge transfer from previous reconstructions.  
As a result, high-quality and low-latency successive imaging for dynamic targets can be achieved.

\section{Imaging-Augmented Communication}
\label{sec-imaging-augmented-communication}

In addition to supporting vertical industry applications through wireless imaging \cite{hu2022metasketch,huang2025cooperative}, this section discusses how the high-quality imaging results provided by INR can be leveraged to enhance communication performance.

\subsection{SE Maximization Problem Formulation}

We assume that the positions of all elements in Fig. \ref{fig-model} remain static.
Consequently, the communication channel $\mathbf{h}_{\text{com}}$ depends only on $\boldsymbol{\omega}$ and is written as $\mathbf{h}_{\text{com}}(\boldsymbol{\omega})$.
The SE, used as the communication performance metric, is defined as
\begin{equation}\label{eq-se1}
{\text{SE}}(\boldsymbol{\omega}) = \log_{2}{\left(1+\frac{P_{\text{t}}\left\|\mathbf{h}_{\text{com}}(\boldsymbol{\omega})\right\|^{2}_2}{\sigma^2_{\text{com}}}\right)},
\end{equation}
where $\sigma^2_{\text{com}}$ denotes the noise variance.
The RIS phase configuration $\boldsymbol{\omega}$ can therefore be optimized to maximize the SE by exploiting environmental information captured through wireless imaging, formulating the following optimization problem:
\begin{equation*}
\begin{aligned}
\text {(P5)}\quad & \max \limits_{\boldsymbol{\omega}} && \left\|\mathbf{h}_{\text{com}}(\boldsymbol{\omega})\right\|^{2}_2 \\
& \text { s.t. } && |\omega_{n_{\text{s}}}|=1, \ \ \forall n_{\text{s}}=1, \ldots, N_\text{s}. \\
\end{aligned}
\end{equation*}
This problem can be solved by the method proposed in \cite{wu2018intelligent}, as detailed in Appendix \ref{appendix-ris-phase-opt}.

\subsection{Scenario 1: Communication User Inside the ROI}
\label{subsec-augment-commun-inside}

In this scenario, we consider a user at the initial access stage who has not established a communication link with the BS, and the user location is unknown priorly \cite{luo2023integrated}.
The user is supposed a human located within the ROI and holding a communication device, such as a mobile phone, as shown in Fig.~\ref{fig-model-in}.
The corresponding channel is denoted as $\mathbf{h}_{\text{com-in}}$, defined in \eqref{eq-commun-channel-inside}.
Accordingly, the RIS phase optimization in (P5) is formulated by substituting   
\begin{equation}\label{eq-commun-ris-optimize-inside}
\mathbf{h}_{\text{com}}(\boldsymbol{\omega}) = \Big( g_{\text{com}}g_{\text{ris}}\mathbf{H}_{\text{tx-ris}}\text{diag}(\mathbf{h}_{\text{ris-ue}}) \Big) \boldsymbol{\omega} + \mathbf{h}_{\text{tx-ue}}.
\end{equation}
To optimize RIS phases for high communication performance, the user location should be obtained.
However, conventional localization methods depend on user cooperation \cite{huang2023joint} and may suffer from low estimation accuracy.
Moreover, prior studies often model the user as a point target, while in practice, the human body position may differ from the communication device position \cite{li2019intelligent,luo2023integrated,zhu2025multi}, potentially degrading the beamforming performance, especially in near-field regions.

To address this, the proposed ISAC system first treats the user as a passive target, performing active sensing to reconstruct an image of the ROI using the proposed INR-based methods.
Hence, the user simultaneously serves as a communication terminal and a target being sensed.
The ROI image is then processed by a hand detection NN to localize the user's hand position \cite{li2019intelligent,ren2016faster}.
Consequently, this detected position is assumed to be the communication device location and is subsequently used for RIS phase optimization, thereby ensuring accurate RIS-aided beamformers and reducing beam scanning time \cite{luo2023integrated}.

\subsection{Scenario 2: Communication User Outside the ROI}
\label{subsec-augment-commun-outside}

In this scenario, we consider a user who has established a communication link with the BS, and the user location can be estimated using RIS-aided localization algorithms \cite{huang2023joint,wang2025reconfigurable}.
The user is assumed to be located outside the ROI, and the communication channel is given by $\mathbf{h}_{\text{com-out}}$, as shown in Fig. \ref{fig-model-out}.
The ROI involves other targets, which introduce additional multipath components to $\mathbf{h}_{\text{com-out}}$, including $\mathbf{h}_{\text{tx-roi-ue}}$, $\mathbf{h}_{\text{tx-roi-ris-ue}}$, and $\mathbf{h}_{\text{tx-ris-roi-ue}}$, as defined in \eqref{eq-commun-channel-outside}.
Although the RIS phases can be optimized for communication without target imaging, the ROI-related multipath components are unknown and cannot be properly considered during RIS phase optimization, potentially reducing the effective energy of $\mathbf{h}_{\text{com-out}}$.

To address this, we propose to employ the INR-based imaging technique to generate a high-quality ROI image, allowing accurate characterization of the ROI-related multipath components.
Based on the channel models in \eqref{eq-commun-channel-single-bounce} and \eqref{eq-commun-channel-twice-bounce}, (P5) can be formulated by substituting \cite{huang2025integrated}
\begin{multline} \label{eq-commun-ris-optimize-outside}
\mathbf{h}_{\text{com}}(\boldsymbol{\omega})  = g_{\text{com}}g_{\text{ris}}\Big(\mathbf{H}_{\text{tx-ris}} \text{diag}(\mathbf{h}_{1}) + \mathbf{H}_{1} \text{diag}(\mathbf{h}_{\text{ris-ue}}) \Big) \boldsymbol{\omega} \\ + \mathbf{h}_{\text{tx-ue}} + \mathbf{h}_{\text{tx-roi-ue}}, 
\end{multline}
where $\mathbf{h}_{1} = \mathbf{h}_{\text{ris-ue}} + \mathbf{H}_{\text{ris-roi}} \text{diag}(\boldsymbol{\sigma}) \mathbf{h}_{\text{roi-ue}}$, and $\mathbf{H}_{1} = \mathbf{H}_{\text{tx-roi}} \text{diag}(\boldsymbol{\sigma}) \mathbf{H}_{\text{roi-ris}}$.
In this manner, the RIS phase design incorporates both ROI- and RIS-related multipaths, leading to improved SE performance.

\section{Numerical Results}
\label{sec-simulation}

\subsection{Experimental Settings}

\subsubsection{Simulation Scenario}

We consider the simulation setup illustrated in Fig. \ref{fig-model}.
The center subcarrier frequency is set to 3 GHz.
The TX and RX are ULAs with $N_{\text{t}}=N_{\text{r}}=8$ antennas, and their center locations are $[10\lambda, 10\lambda, 0]^{\text{T}}$ and $[10\lambda, -10\lambda, 0]^{\text{T}}$, respectively.
The integrated antenna gains, i.e., $G_{\text{com}}$ and $G_{\text{sen}}$, are set to 4 \cite{sharma2022mimo}.
The received noise power at each antenna of the RX is set to $P_{\text{n}}=-110$ dBm \cite{li2023toward}.
The RIS is deployed in the yOz plane with its center located at $[0, 0, 0]^{\text{T}}$, consisting of $50\times50$ elements with size $\xi_{\text{s}} = \lambda/2$, resulting in a uniform planar array of side length 2.5~m.
We consider a human imaging scenario, in which the ROI is typically assumed to be 2D in literature \cite{wang2025dreamer,hu2022metasketch,li2019intelligent}, whereas 3D imaging results are presented in Sec. \ref{subsec-result-threeD}.
The center of the ROI is located at $[D, 0, 0]^{\text{T}}$ with dimensions of $2\text{m}\times 2\text{m}$.
The ROI lies in the yOz plane and is discretized into $100\times100$ pixels, with each pixel sized $\lambda/5 \times \lambda/5$, which is smaller than the diffraction resolution limit presented in \cite{huang2024fourier}.

\subsubsection{Data Generation}

\begin{figure}[t]
    \centering
    \includegraphics[width=0.85\linewidth]{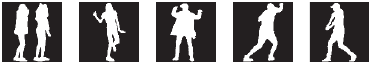}
    \captionsetup{font=footnotesize}
    \caption{Example images of the synthesized dataset.}
    \label{fig-dataset}
\end{figure}

We employ a human segmentation dataset derived from TikTok dance videos on Kaggle \cite{kaggle2023dataset} to simulate human targets in the ROI.
A total of 2,615 segmented images are extracted, cropped, resized, and converted into grayscale images of size $100 \times 100$. Pixel values are normalized to the range $[0, 4\pi A^2/\lambda^2]$, representing the radar cross section (RCS) of a voxel with area $A=\lambda/5\times\lambda/5$ \cite{huang2024ris}.
These normalized pixel values reflect the scattering characteristics of the corresponding voxels.
Examples from the synthesized dataset are shown in Fig.~\ref{fig-dataset}.
Note that the GT images are not required by the proposed INR-based imaging method; they are used only to generate CSI and evaluate imaging accuracy in the experiments.
The CSI is generated using the ray tracing channel model given in \cite{goldsmith2005wireless}, as presented in Sec. \ref{sec-system-model}.

\subsubsection{Training Details}

The NN is trained for 5000 epochs with an initial learning rate of $10^{-3}$.
The learning rate is reduced by 50\% if the validation accuracy does not improve for 50 consecutive epochs, and training stops if no improvement is observed over 200 epochs.
The positional encoding dimension is set to $N_{\text{pe}}=256$, and the $\ell_{2}$-norm of the NN gradients is clipped to be below 2.
Training is performed using the Adam optimizer on an Nvidia 4090 GPU with the PyTorch platform.
Unless otherwise specified, the training grid $\boldsymbol{\mathcal{G}}$ is set equal to the output image grid $\boldsymbol{\mathcal{G}}^{\circ}$.

\subsubsection{Performance Evaluation Metrics}

The following metrics are used to evaluate imaging quality:

\textbf{(1) Mean Square Error (MSE):} Measures the per-voxel error between the predicted $\hat{\boldsymbol{\sigma}}$ and GT $\boldsymbol{\sigma}$:
\begin{equation}\label{eq-mse}
\text{MSE}=\|\hat{\boldsymbol{\sigma}}-\boldsymbol{\sigma}\|^2_2/N_{\text{v}}.
\end{equation}

\textbf{(2) Peak Signal-to-Noise Ratio (PSNR):} Assesses the ratio of the peak signal power to noise:
\begin{equation}\label{eq-psnr}
\text{PSNR}=10\log_{10}\left(\sigma_{\max}^2/\text{MSE}\right),
\end{equation}
where $\sigma_{\max}$ is the maximum voxel value.
MSE and PSNR jointly measure pixel-level accuracy.

\begin{table*}[t]
    \renewcommand{\arraystretch}{1.4}
    \centering
    \fontsize{8}{8}\selectfont
    \captionsetup{font=small}
    \captionof{table}{Imaging results with different system and training settings (unit for time: second).}\label{tab-result-compare1}
    \begin{threeparttable}
        \begin{tabular}{rcccccc}
            \specialrule{1pt}{0pt}{-1pt}\xrowht{10pt}
            RIS phase & \textbf{no RIS} & \textbf{DFT} & random & random & random & random \\
            Activation function & sin & sin & \textbf{ReLU} & sin & sin & sin \\
            Positional encoding & \checked & \checked & \checked & \faTimes & \checked & \checked \\
            $\alpha$ & 0 & 0 & 0 & 0 & \textbf{0.01} & 0 \\
            \hline
            MSE & 0.2440 & 0.0019 & 0.2485 & 0.0018 & 0.0012 & \textbf{0.0011} \\
            PSNR (dB) & 6.1258 & 27.1173 & 6.0472 & 27.4085 & 29.2841 & \textbf{29.5889} \\
            SSIM & 0.0625 & 0.9228 & 0.1296 & 0.9427 & \textbf{0.9889} & 0.9755 \\
            \hline
            \specialrule{0pt}{1pt}{0pt} Train time (200 epochs) & \textbf{0.9427} & 1.0836 & 1.0057 & 1.0409 & 1.0356 & 1.1073 \\
            \specialrule{0pt}{1pt}{0pt} Imaging result (200 epochs) & 
            \adjustbox{valign=m}{\includegraphics[width=1cm]{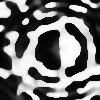}} & 
            \adjustbox{valign=m}{\includegraphics[width=1cm]{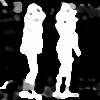}} & 
            \adjustbox{valign=m}{\includegraphics[width=1cm]{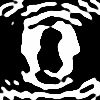}} & 
            \adjustbox{valign=m}{\includegraphics[width=1cm]{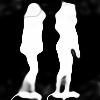}} & 
            \adjustbox{valign=m}{\includegraphics[width=1cm]{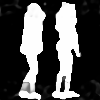}} & 
            \adjustbox{valign=m}{\includegraphics[width=1cm]{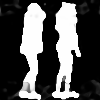}} \\
            \specialrule{0pt}{1pt}{0pt} Train time (5,000 epochs) & \textbf{23.5677} & 27.0909 & / & 26.0224 & 25.8896 & 27.6816 \\
            \specialrule{0pt}{1pt}{0pt} Imaging result (5,000 epochs) & 
            \adjustbox{valign=m}{\includegraphics[width=1cm]{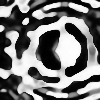}} & 
            \adjustbox{valign=m}{\includegraphics[width=1cm]{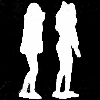}} & 
            \adjustbox{valign=m}{\includegraphics[width=1cm]{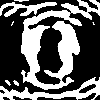}} & 
            \adjustbox{valign=m}{\includegraphics[width=1cm]{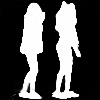}} & 
            \adjustbox{valign=m}{\includegraphics[width=1cm]{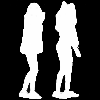}} & 
            \adjustbox{valign=m}{\includegraphics[width=1cm]{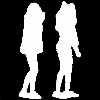}} \\
            \specialrule{1pt}{1pt}{0pt}
        \end{tabular}
    \end{threeparttable}
\end{table*}

\textbf{(3) Structural Similarity Index Measure (SSIM):} Evaluates the structural similarity between $\hat{\boldsymbol{\sigma}}$ and $\boldsymbol{\sigma}$ \cite{wang2025dreamer}:
\begin{equation}\label{eq-ssim}
\text{SSIM}=\frac{\left(2 \mu_{\boldsymbol{\sigma}} \mu_{\hat{\boldsymbol{\sigma}}}+c_{1}\right)\left(2 \theta_{\boldsymbol{\sigma} \hat{\boldsymbol{\sigma}}}+c_{2}\right)}{\left(\mu_{\boldsymbol{\sigma}}^{2}+\mu_{\hat{\boldsymbol{\sigma}}}^{2}+c_{1}\right)\left(\theta_{\boldsymbol{\sigma}}^{2}+\theta_{\hat{\boldsymbol{\sigma}}}^{2}+c_{2}\right)},
\end{equation}
where $\mu_{\boldsymbol{\sigma}}$ ($\mu_{\hat{\boldsymbol{\sigma}}}$) and $\theta_{\boldsymbol{\sigma}}^{2}$ ($\theta_{\hat{\boldsymbol{\sigma}}}^{2}$) are the mean and variance of $\boldsymbol{\sigma}$ ($\hat{\boldsymbol{\sigma}}$), respectively, and
$\theta_{\boldsymbol{\sigma} \hat{\boldsymbol{\sigma}}}$ is the covariance between them.
Constants $c_1$ and $c_2$ follow the default settings in the \texttt{skimage} python package.
SSIM ranges from 0 to 1, with higher values indicating better visual similarity.
Unlike MSE and PSNR, SSIM captures structural and perceptual fidelity, reflecting human visual perception.

\subsection{Results and Discussions}

\subsubsection{Imaging Results with Different Systems and Training Settings}

\begin{figure}[t]
    \centering
    \includegraphics[width=0.8\linewidth]{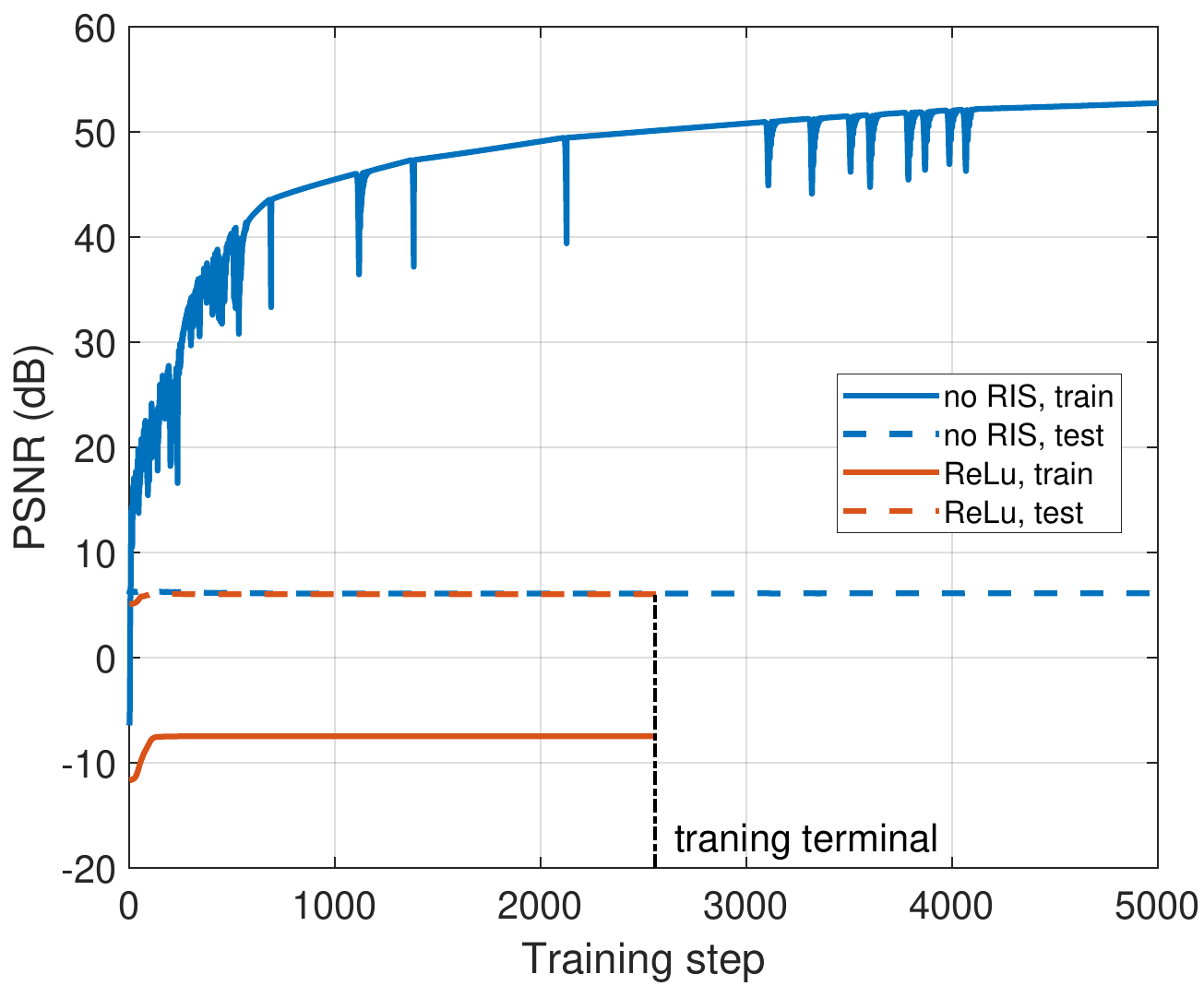}
    \captionsetup{font=footnotesize}
    \caption{Train and test PSNR during the training process for ``no RIS'' and ``ReLU'' scenarios.}
    \label{fig-result-train-process-no-ris}
\end{figure}

We evaluate the effectiveness of the proposed INR-based imaging algorithm under $K=40$ distinct RIS configurations, with the ROI positioned at $D=40\lambda$.
Table \ref{tab-result-compare1} shows that the proposed method performs well, especially when the sinusoidal activation function and positional encoding are used.
After 5000 training epochs, the model produces highly accurate images, with SSIM values approaching 0.99.
Even at 200 epochs, the INR already generates nearly perfect images, with the cost of time in approximately one second.
In contrast, using the ReLU activation function results in severely distorted images, and omitting positional encoding also degrades PSNR and SSIM performance.
Random RIS phase configurations outperform the DFT codebook in sensing performance.
Although omitting the RIS simplifies the forward model and slightly speeds up training, the resulting images are ineffective, indicating the importance of diverse CSI measurements enabled by RIS phase control.

Fig. \ref{fig-result-train-process-no-ris} illustrates the training and testing PSNRs for the ``no RIS'' and ``ReLU'' configurations.
Without RIS, training PSNR increases, but testing PSNR remains below 10 dB, implying overfitting and poor generalization due to insufficient image information.
For the ReLU-based NN, neither training nor testing PSNR improves, indicating that ReLU fails to capture the high-frequency features essential for image reconstruction, despite having access to sufficient CSI data.

\subsubsection{Hyperparameter Discussions}

\begin{table}[t]
    \renewcommand{\arraystretch}{1.4}
    \centering
    \fontsize{8}{8}\selectfont
    \captionsetup{font=small}
    \captionof{table}{Imaging performance versus $\alpha$.}\label{tab-result-alpha}
    \begin{threeparttable}
        \begin{tabular}{cccccc}
            \specialrule{1pt}{0pt}{-1pt}\xrowht{10pt}
            $\alpha$ & $10^0$ & $10^{-0.5}$ & $10^{-1.0}$ & $10^{-1.5}$ & $10^{-2.0}$ \\
            \specialrule{0.5pt}{1pt}{1pt}
            PSNR (dB) & 16.6162 & 21.2241 & 24.8158 & 27.1059 & \textbf{29.2841} \\
            SSIM & 0.7514 & 0.8756 & 0.9513 & 0.9807 & \textbf{0.9889} \\
            \specialrule{1pt}{1pt}{0pt}
        \end{tabular}
    \end{threeparttable}
\end{table}

\begin{figure}[t]
    \centering
    \includegraphics[width=0.8\linewidth]{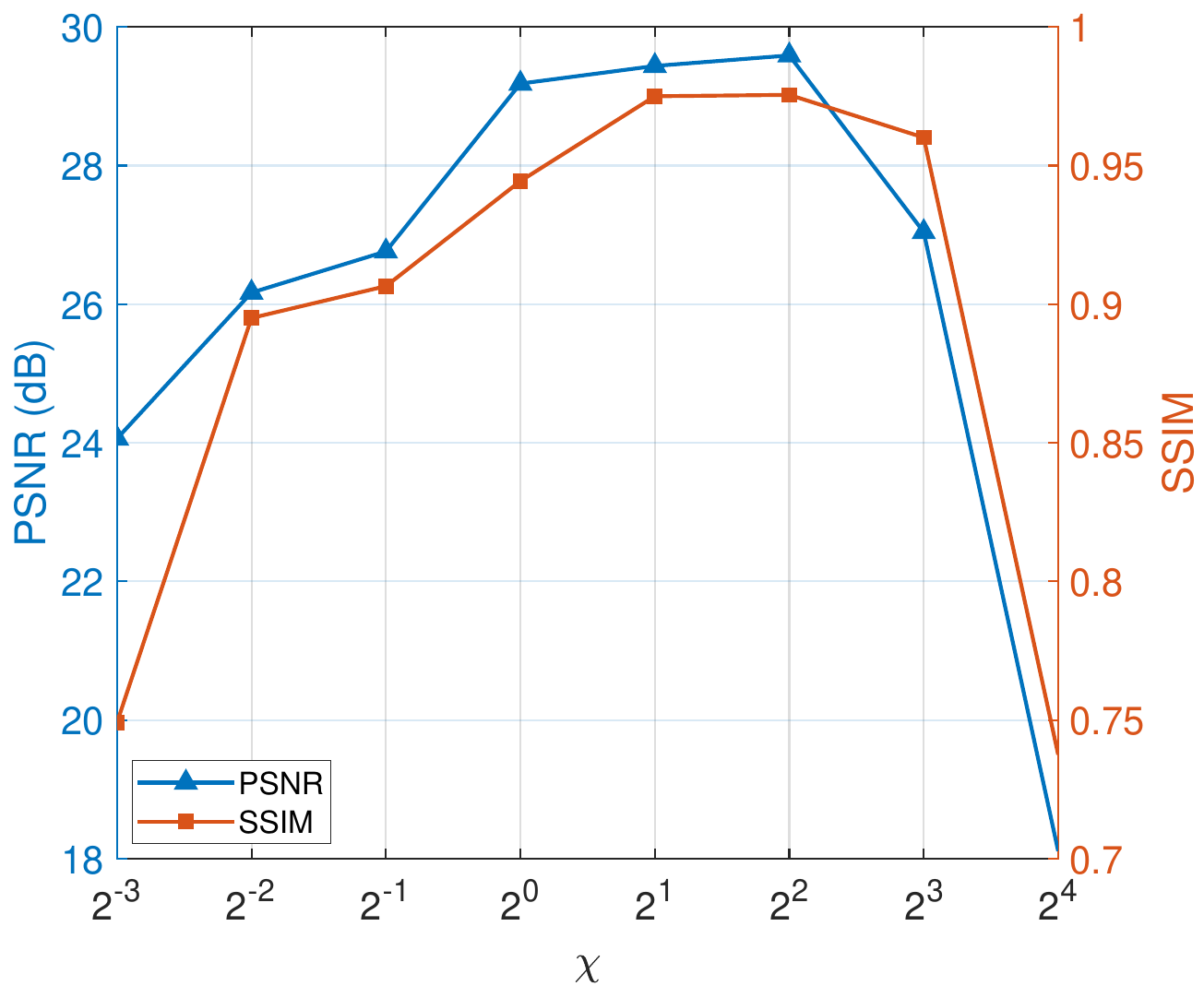}
    \captionsetup{font=footnotesize}
    \caption{Imaging performance with varying $\chi$.}
    \label{fig-result-pe}
\end{figure}

This subsection discusses the choice of the hyperparameter $\alpha$ in the loss function and $\chi$ in positional encoding.

First, we have employed a sparsity penalty $\alpha$ in the loss function to stabilize the NN training process and prevent meaningless results.
To address the influences of $\alpha$, additional simulations are conducted, and the results are listed in Table \ref{tab-result-alpha}.
When $\alpha$ is large, the sparsity penalty term dominates in the loss function, leading to low supervision ability of the physical model.
Thus, the imaging performance is degraded.
According to Table \ref{tab-result-compare1}, $\alpha=0$ and $\alpha=0.01$ derive approximately the same imaging performance.
Consequently, $\alpha=0.01$ is a desirable choice for NN training and has been utilized for subsequent simulations.

Second, the variance of the elements in matrix $\mathbf{B}$ is determined by $\chi^2$, influencing Fourier frequencies in positional encoding.
According to \cite{tancik2020fourier}, $\chi$ is chosen for each task with a hyperparameter sweep, and our simulation results with respect to $\chi$ are shown in Fig. \ref{fig-result-pe}.
In our considered scenarios, $\chi=4$ realizes the best performance of PSNR and SSIM.
Thus, this value is employed for the following subsections.

\subsubsection{Imaging Result Comparison of INR with FT and CS Algorithms}

\begin{table}[t]
    \renewcommand{\arraystretch}{1.4}
    \centering
    \fontsize{8}{8}\selectfont
    \captionsetup{font=small}
    \captionof{table}{Imaging results using different algorithms.}\label{tab-result-compare-ft-cs}
    \begin{threeparttable}
        \begin{tabular}{p{1.4cm}<{\centering}ccccc}
            \specialrule{1pt}{0pt}{-1pt}\xrowht{10pt}
            Algorithm & GT & INR & INR & FT & CS \\
            $K$ & / & 500 & 2500 & 2500 & 2500 \\
            \specialrule{0.5pt}{1pt}{1pt} $\xi_{\text{v}}=\lambda/5$ & 
            \adjustbox{valign=m}{\includegraphics[width=1cm]{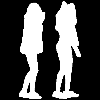}} & 
            \adjustbox{valign=m}{\includegraphics[width=1cm]{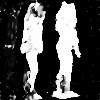}} & 
            \adjustbox{valign=m}{\includegraphics[width=1cm]{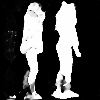}} & 
            \adjustbox{valign=m}{\includegraphics[width=1cm]{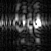}} & 
            \adjustbox{valign=m}{\includegraphics[width=1cm]{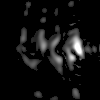}} \\
            \specialrule{0pt}{1pt}{1pt} $\xi_{\text{v}}=\lambda/2.5$ & 
            \adjustbox{valign=m}{\includegraphics[width=1cm]{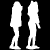}} & 
            \adjustbox{valign=m}{\includegraphics[width=1cm]{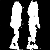}} & 
            \adjustbox{valign=m}{\includegraphics[width=1cm]{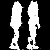}} & 
            \adjustbox{valign=m}{\includegraphics[width=1cm]{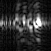}} & 
            \adjustbox{valign=m}{\includegraphics[width=1cm]{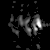}} \\
            \specialrule{0pt}{1pt}{1pt} $\xi_{\text{v}}=\lambda$ & 
            \adjustbox{valign=m}{\includegraphics[width=1cm]{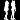}} & 
            \adjustbox{valign=m}{\includegraphics[width=1cm]{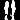}} & 
            \adjustbox{valign=m}{\includegraphics[width=1cm]{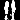}} & 
            \adjustbox{valign=m}{\includegraphics[width=1cm]{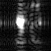}} & 
            \adjustbox{valign=m}{\includegraphics[width=1cm]{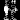}} \\
            \specialrule{1pt}{1pt}{0pt}
        \end{tabular}
    \end{threeparttable}
\end{table}

We compare the imaging performance of FT \cite{huang2024fourier}, CS \cite{huang2024ris,hu2022metasketch,tong2025computational}, and proposed INR algorithms.
To maintain consistent simulation settings, only the TX-ROI-RIS-RX path is employed for image reconstruction, the numbers of antennas at the TX and RX are both set to 1, and the RIS phases are configured using a DFT codebook.
The imaging distance is set to $D = 40\lambda$.
As shown in Table~\ref{tab-result-compare-ft-cs}, the proposed INR-based imager significantly outperforms traditional FT and CS methods.
With only $K = 500$ RIS phase variations, INR achieves high imaging quality.
In contrast, even with $K = 2500$ measurements, FT and CS generate degraded images due to strong pixel-wise channel correlations, which limit their ability to resolve fine details.
This highlights the advantage of leveraging DL techniques, as INR is capable of learning intricate image features directly from CSI data.
Furthermore, INR enables the generation of super-resolution images with pixel size $\xi_{\text{v}}$ much smaller than the resolution limit established in \cite{huang2024fourier}.
By comparison, FT and CS can only produce coarse approximations when $\xi_{\text{v}} = \lambda$.
Therefore, the INR framework presents an effective alternative to conventional model-based imaging approaches, offering enhanced resolution and robustness in underdetermined sensing scenarios.

\subsubsection{Imaging Performance versus Distance and RIS Phase Variation Number} 

\begin{figure}
    \centering
    \includegraphics[width=0.8\linewidth]{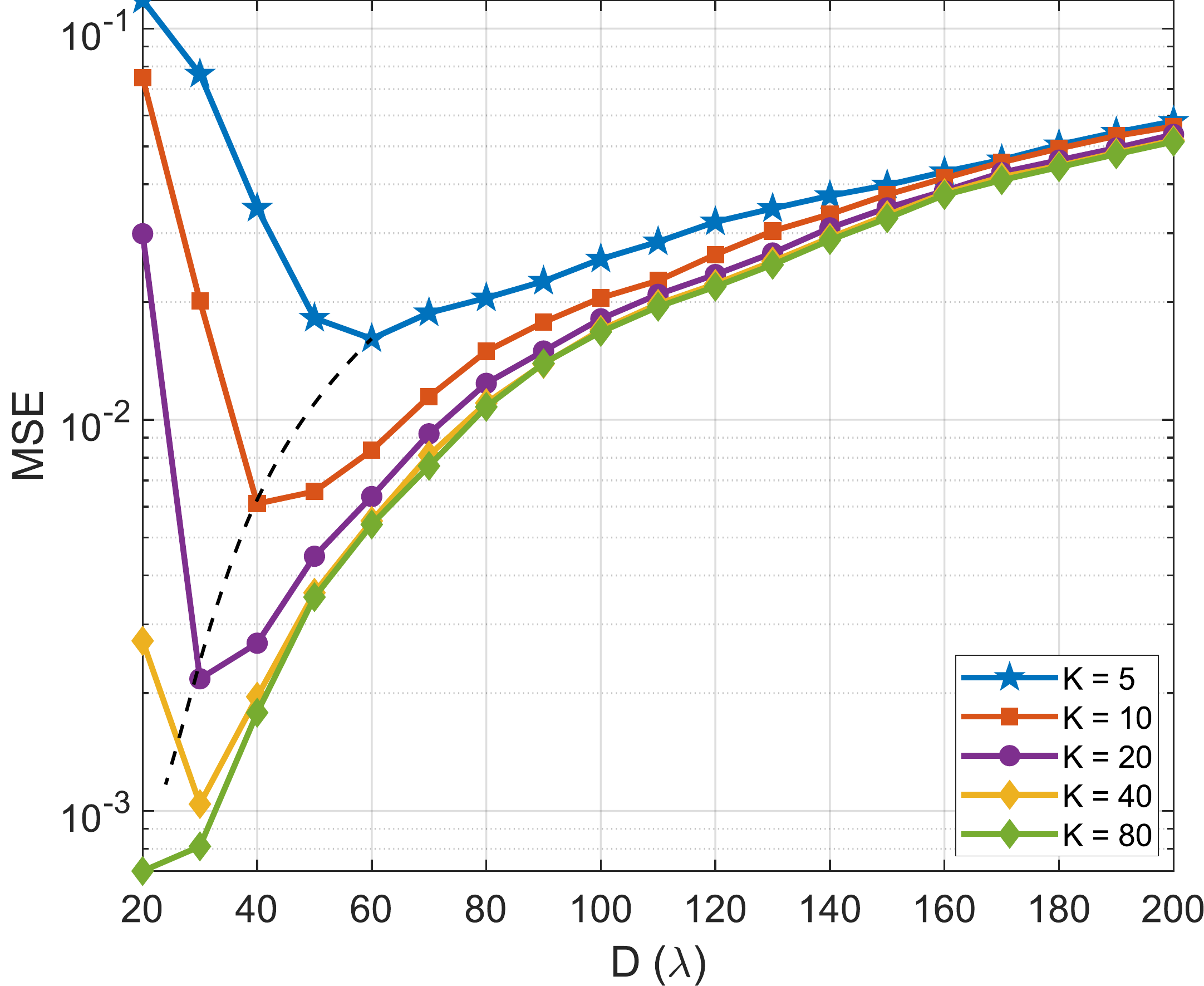}
    \captionsetup{font=footnotesize}
    \caption{MSE versus imaging distance and RIS phase variation number.}
    \label{fig-result-distance-glimpse-mse}
\end{figure}

\begin{figure}
    \centering
    \includegraphics[width=0.8\linewidth]{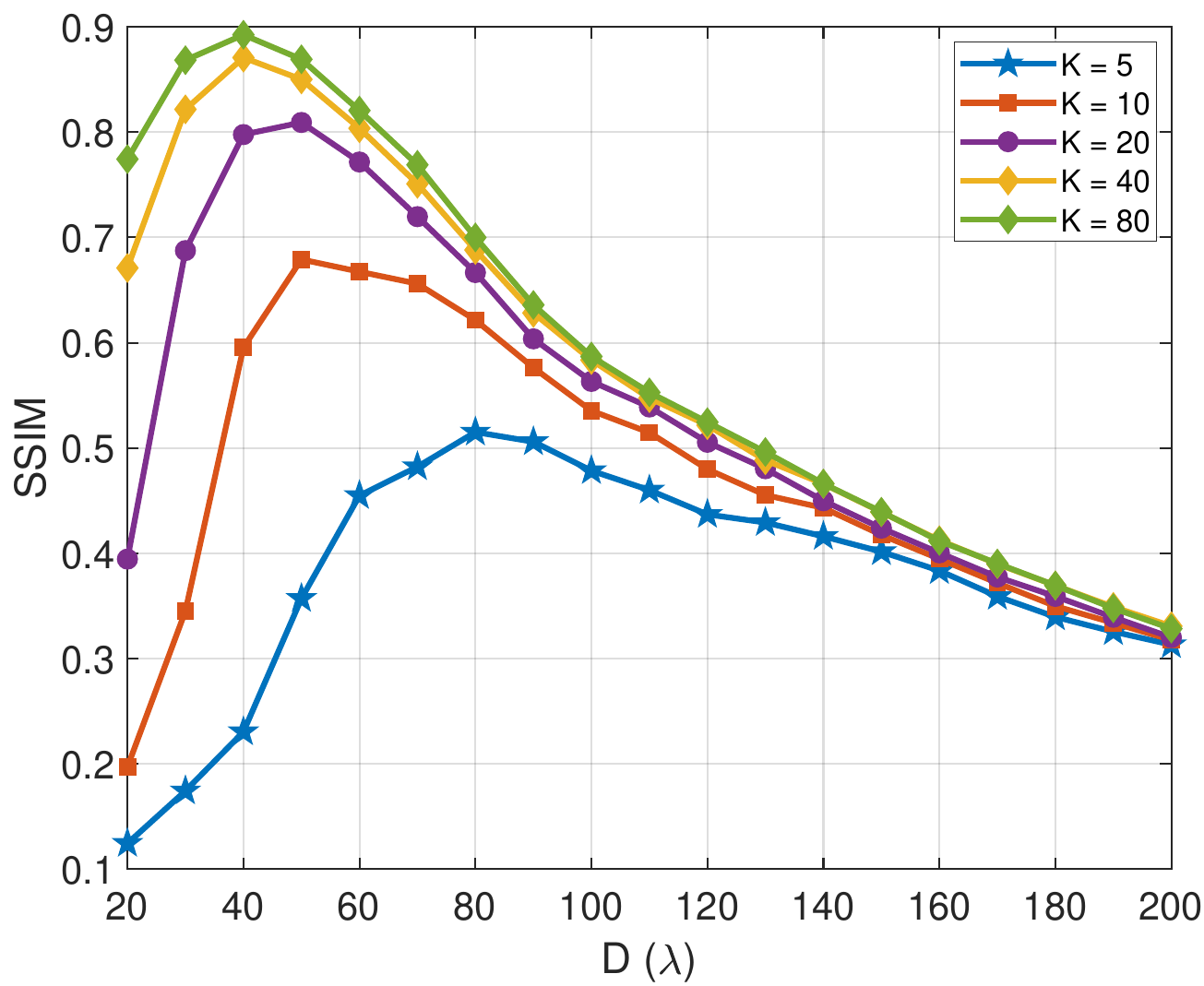}
    \captionsetup{font=footnotesize}
    \caption{SSIM versus imaging distance and RIS phase variation number.} 
    \label{fig-result-distance-glimpse-ssim}
\end{figure}

This subsection investigates the effects of imaging distance $D$ and the number of RIS phase variations $K$ on the imaging quality.
To reduce simulation time, the maximum training epoch is set to 1,000.
The average MSE and SSIM results for 1,000 images using the proposed INR algorithm are shown in Fig.~\ref{fig-result-distance-glimpse-mse} and Fig.~\ref{fig-result-distance-glimpse-ssim}, respectively.
Several key insights can be drawn from these results.

\textit{First}, both MSE and SSIM indicate that imaging quality initially improves and then degrades as the distance $D$ increases from $20\lambda$ to $200\lambda$.
This suggests the existence of an optimal imaging distance $D_0$, analogous to the focal distance in optical imaging, where performance is maximized.
However, $D_0$ varies with $K$; for instance, $D_0 = 60\lambda$ when $K = 5$, and $D_0 = 30\lambda$ when $K = 40$.
In general, $D_0$ decreases as $K$ increases, as illustrated by the dotted black line in Fig.~\ref{fig-result-distance-glimpse-mse}.
Therefore, collecting more CSI measurements allows the proposed INR-based imager to achieve optimal performance at shorter distances.
These findings differ from traditional results \cite{huang2024ris,huang2024fourier}, which typically conclude that imaging performance deteriorates monotonically with increasing $D$.

\textit{Second}, when $D < D_0$, imaging performance is primarily constrained by the number of RIS phase variations $K$.
In this region, the short imaging distance makes the ROI generate a large field of view to the RIS aperture, requesting a large $K$ to capture adequate information about the ROI, which resembles the scenario when a man observes a near and large object.
As a result, a low $K$ leads to insufficient sensing perspectives and degraded performance, similar to the no-RIS scenario in Table~\ref{tab-result-compare1}.
Increasing $K$ from 5 to 80 can enrich CSI measurements and ROI-related information, reducing MSE from around $10^{-1}$ to $10^{-3}$ and increasing SSIM from 0.1 to 0.8 at $D = 20\lambda$.

\textit{Third}, when $D > D_0$, the influence of $K$ diminishes.
For example, when $D$ approaches $200\lambda$, the MSE and SSIM values for $K = 5$ and $K = 80$ become nearly identical.
This indicates that the performance bottleneck shifts from RIS phase diversity to the increasing channel correlations between the RIS and ROI.
These correlations make it more difficult to distinguish adjacent pixels as $D$ increases, consistent with analysis in \cite{huang2024ris,huang2024fourier}.
Consequently, imaging resolution deteriorates at larger distances. 
This condition also shares similarities with optical imaging systems where multiple observations may not significantly improve the resolution at large imaging distances.

\subsubsection{Imaging Performance versus Noise Power and TX Antenna Number}

\begin{figure}
    \centering
    \includegraphics[width=0.8\linewidth]{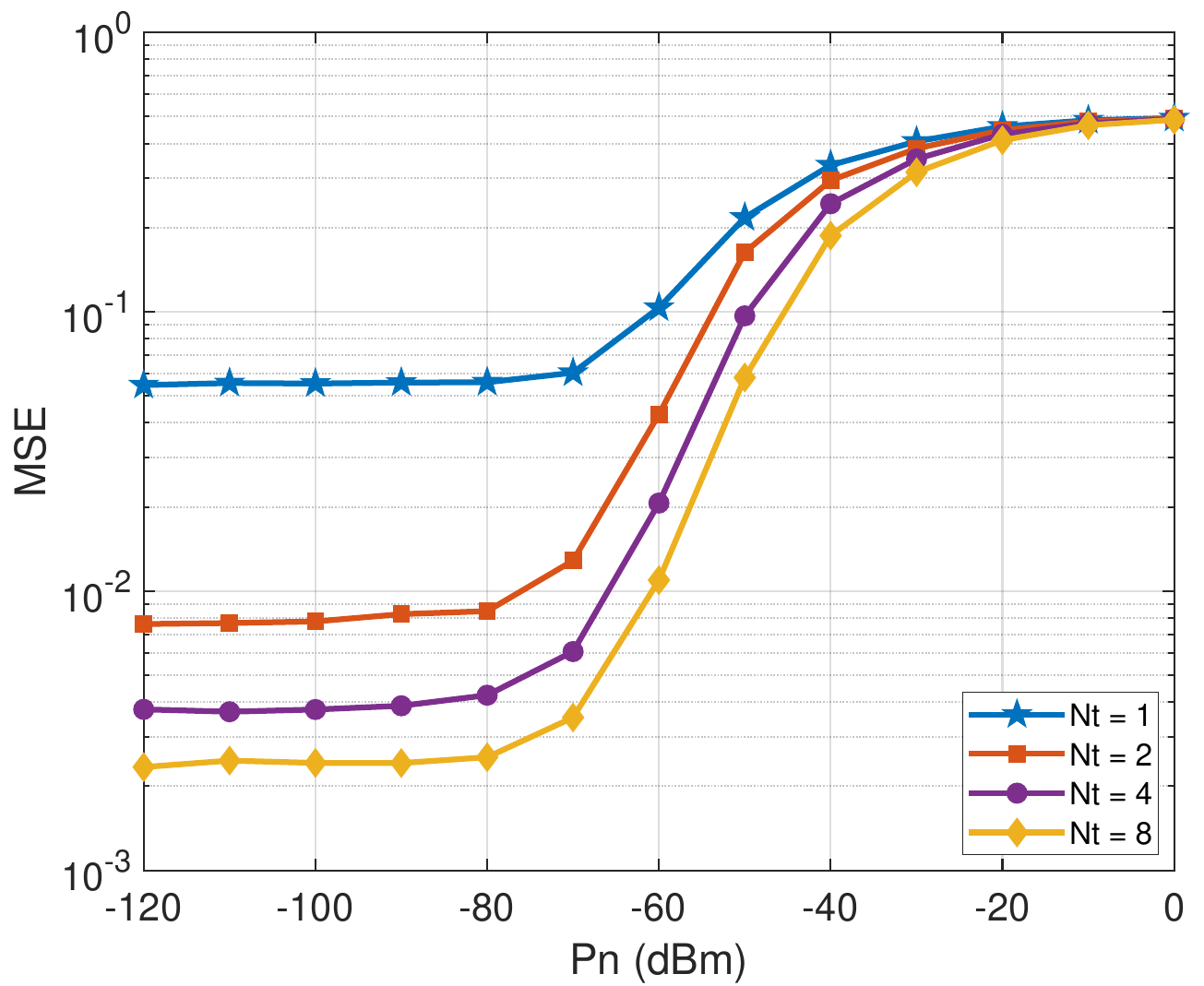}
    \captionsetup{font=footnotesize}
    \caption{MSE versus noise power and TX antenna number.} 
    \label{fig-result-noise-antenna-mse}
\end{figure}

\begin{figure}
    \centering
    \includegraphics[width=0.8\linewidth]{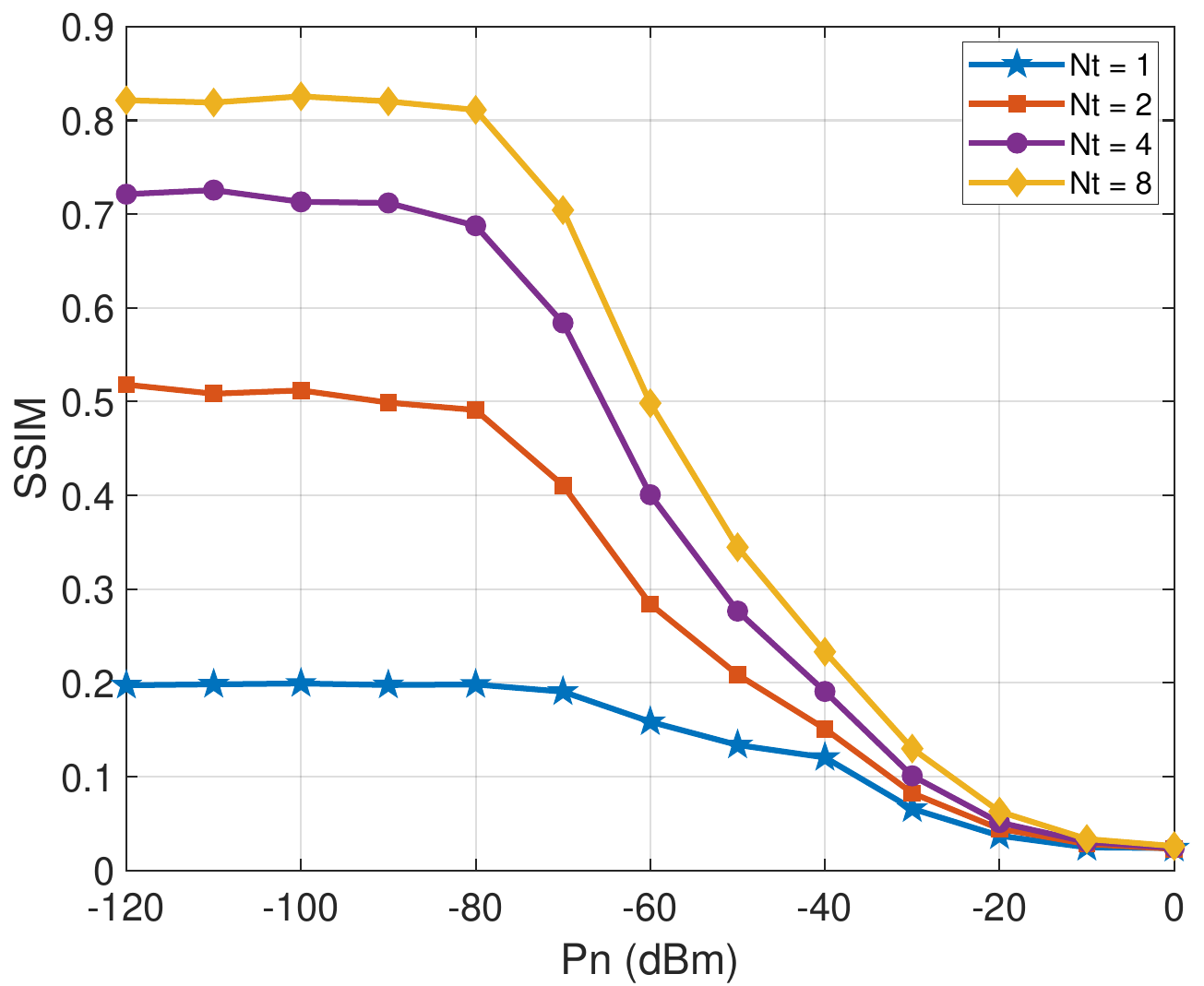}
    \captionsetup{font=footnotesize}
    \caption{SSIM versus noise power and TX antenna number.} 
    \label{fig-result-noise-antenna-ssim}
\end{figure}

Fig.~\ref{fig-result-noise-antenna-mse} and Fig.~\ref{fig-result-noise-antenna-ssim} illustrate the effects of varying noise power $P_{\text{n}}$ and the number of TX antennas $N_{\text{t}}$ on imaging performance, where $P_{\text{n}}$ may originates from channel estimation and background calibration errors.
The number of RX antennas is set to $N_{\text{r}} = 4$, and the transmit power is fixed at $P_{\text{t}} = 20$ dBm.
Results are averaged over 1,000 test images, using $K = 40$ RIS phase variations and an imaging distance of $D = 40\lambda$.
The results show that the proposed algorithm is robust to noise when $P_{\text{n}} < -80$ dBm, with minimal impacts from additive noise.
However, when $P_{\text{n}} > -80$ dBm, the imaging quality degrades as noise increases, leading to noise-dominated results as $P_{\text{n}}$ approaches 0 dBm.
Under fixed transmit power, increasing the number of TX antennas improves performance in both MSE and SSIM metrics, despite the reduced per-antenna signal power.
These improvements are attributed to the enhanced spatial resolution offered by the TX antenna array.
Nevertheless, the performance gain diminishes as $N_{\text{t}}$ becomes large, indicating a trade-off between hardware cost and imaging quality.

\subsubsection{Imaging Performance under TX Position Errors}
\label{subsec-result-TXerror}

The calculation of the physical model relies on the locations of the TX, RX, and RIS, which are typically measured manually and are thus susceptible to measurement errors.
We have conducted simulations to evaluate the imaging performance against TX location accuracy ($\Delta$), and the results are shown in Table \ref{tab-result-TX-error}.
When $\Delta=10^{-2}\lambda$, the imaging result is comparable to that with no location errors.
However, as $\Delta$ increases, the imaging results become progressively distorted.
At $\Delta=10^{-0.5}\lambda$, significant noise artifacts appear, potentially obscuring the human body images.
Consequently, high-precision antenna localization is essential for ensuring high-quality image generation.
This vulnerability to location errors is also a common challenge for traditional imaging algorithms.

\begin{table}[t]
    \renewcommand{\arraystretch}{1.4}
    \centering
    \fontsize{8}{8}\selectfont
    \captionsetup{font=small}
    \captionof{table}{Imaging results with different TX position errors.}\label{tab-result-TX-error}
    \begin{threeparttable}
        \begin{tabular}{cccccc}
            \specialrule{1pt}{0pt}{-1pt}\xrowht{10pt}
            $\Delta$ ($\lambda$) & 0 & $10^{-2}$ & $10^{-1.5}$ & $10^{-1}$ & $10^{-0.5}$ \\
            \specialrule{0.5pt}{1pt}{1pt}
            Result & \adjustbox{valign=m}{\includegraphics[width=1cm]{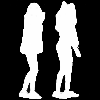}} & 
            \adjustbox{valign=m}{\includegraphics[width=1cm]{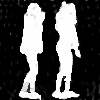}} & 
            \adjustbox{valign=m}{\includegraphics[width=1cm]{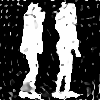}} & 
            \adjustbox{valign=m}{\includegraphics[width=1cm]{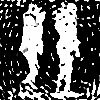}} &
            \adjustbox{valign=m}{\includegraphics[width=1cm]{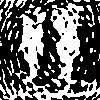}} \\
            \specialrule{1pt}{1pt}{0pt}
        \end{tabular}
    \end{threeparttable}
\end{table}

\subsubsection{INR-Based 3D Imaging}
\label{subsec-result-threeD}

\begin{figure}
    \centering
    \includegraphics[width=\linewidth]{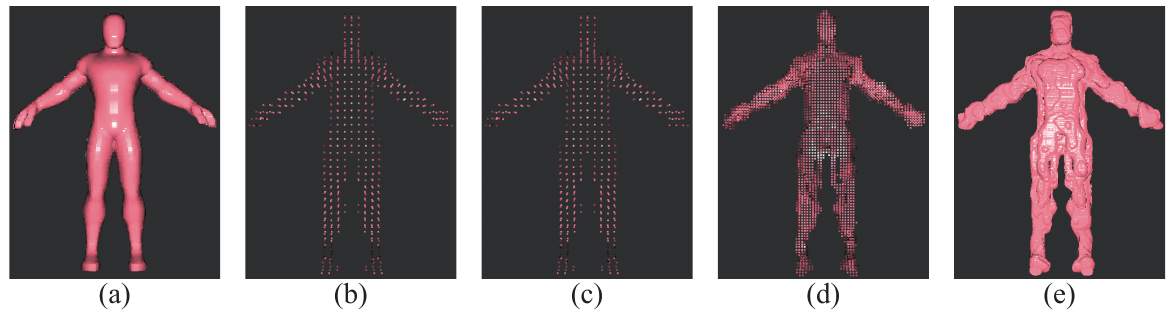}
    \captionsetup{font=footnotesize}
    \caption{3D imaging results: (a) Original 3D image; (b) Discretized GT 3D image for CSI generation; (c) Learned 3D image with coarse resolutions; (d) Rendered image with the resolution of $\lambda/5$; (e) Rendered image with the resolution of $\lambda/20$.}
    \label{fig-threeD}
\end{figure}

We further evaluate the 3D imaging ability of the proposed INR-based imaging method using a 3D human body model shown in Fig. \ref{fig-threeD}(a).
3D imaging has significantly enlarged the size of the ROI, leading to a large image dimension $N_\text{v}$ and high computational complexity of the physical model.
To accelerate image formulation, we employ a coarse grid for NN training with a voxel size of $\lambda/2$, discretizing the ROI of $1.6\times 0.4\times 2\ \text{m}^3$ into a point cloud image with $32\times 8\times 40$ voxels, as depicted in Fig. \ref{fig-threeD}(b).
Furthermore, the RIS size is set to $100\times 100$, and 10 subcarriers averagely spaced on a bandwidth of 400 MHz are employed.
The training results are shown in Figs. \ref{fig-threeD}(c-e), where Fig. \ref{fig-threeD}(c) uses the same coarse grid for NN training, showing high consistency with Fig. \ref{fig-threeD}(b) but low visualizability.
Since the discretized training grid in Fig. \ref{fig-threeD}(b) has lost substantial image details, the NN may not learn all the information involved in the original 3D model.
However, the proposed INR-based imaging method can render images with arbitrary resolutions, and off-grid voxel values can be obtained by increasing the sampling grid density.
Consequently, we can derive Fig. \ref{fig-threeD}(d) and Fig. \ref{fig-threeD}(e) with resolutions of $\lambda/5$ and $\lambda/20$, respectively.
These images have presented smoother surfaces and more semantic details.
Therefore, even trained with coarse grids, the proposed INR-based imaging method can still generate high-resolution images, which is significantly superior to traditional model-based and data-driven DL methods.

\subsubsection{Imaging Performance under Multipath Interference}

\begin{figure*}[t]
\centering
\captionsetup{font=footnotesize}
\begin{subfigure}[b]{0.3\linewidth}
\centering
\includegraphics[width=\linewidth]{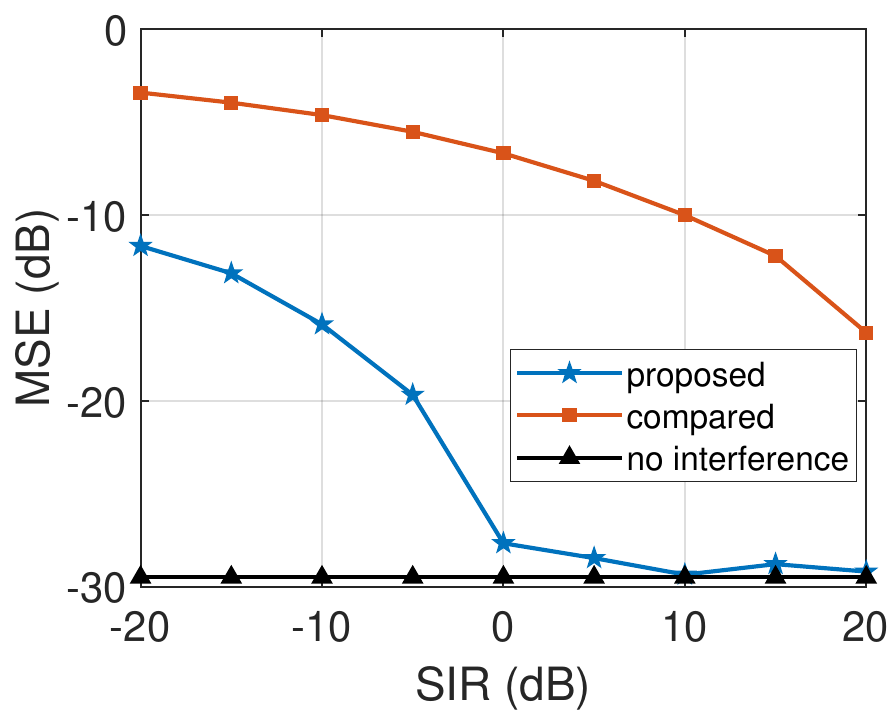}
\caption{}
\label{fig-result-interference-mse}
\end{subfigure}
\begin{subfigure}[b]{0.3\linewidth}
\centering
\includegraphics[width=\linewidth]{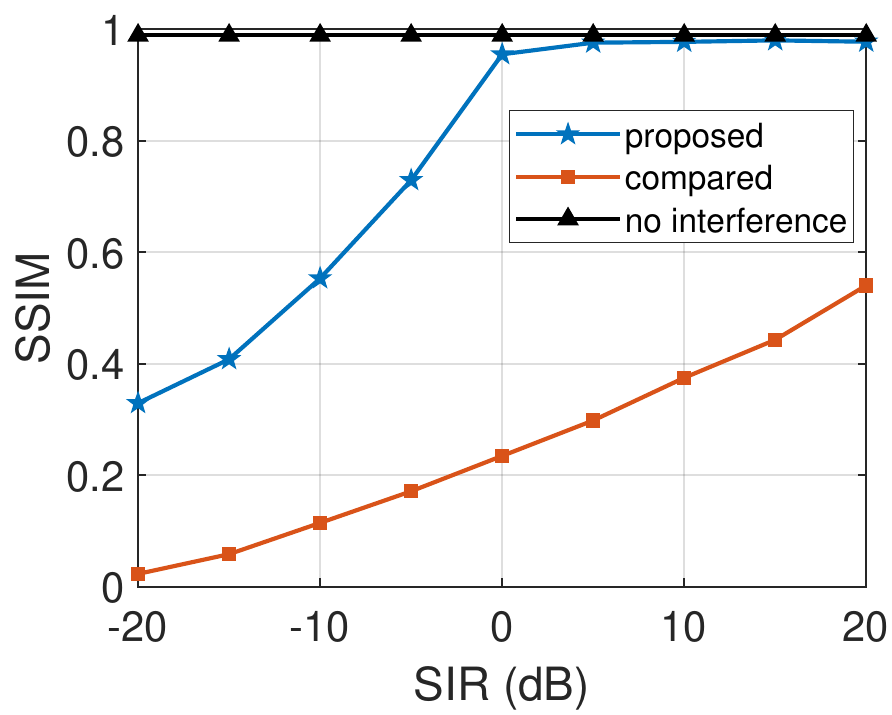}
\caption{}
\label{fig-result-interference-ssim}
\end{subfigure}
\begin{subfigure}[b]{0.3\linewidth}
\centering
\includegraphics[width=\linewidth]{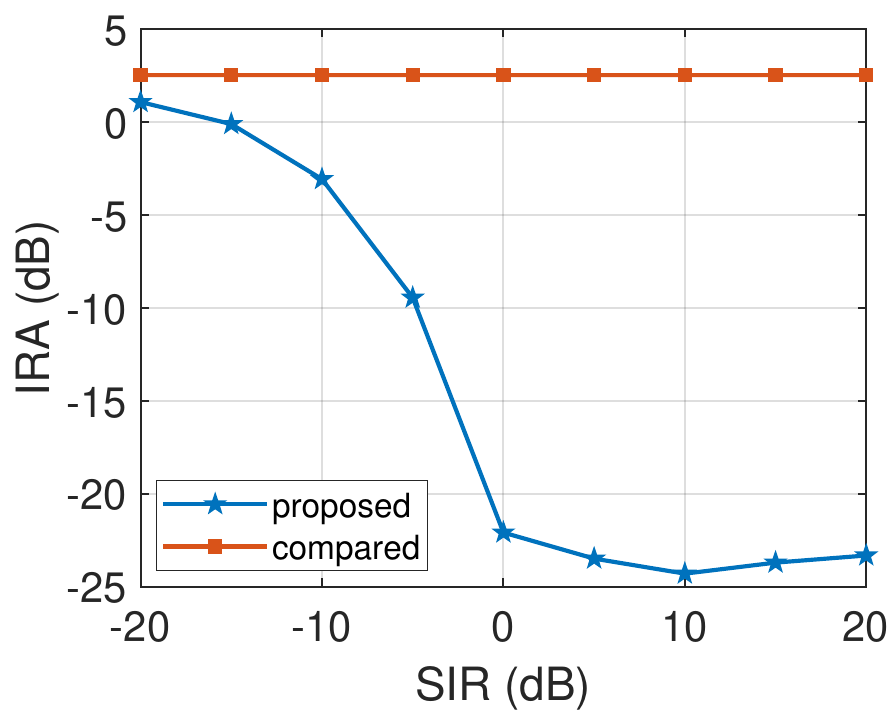}
\caption{}
\label{fig-result-interference-ira}
\end{subfigure}

\caption{Performances of target imaging under multipath interference ${\mathbf{H}}_{\text{ima}, k}$ in \eqref{eq-joint-learning-channel}. ``proposed'' denotes the results obtained using the joint imaging and interference learning method in Sec.~\ref{subsubsec-joint-multipath-learning}, ``compared'' refers the results conducting imaging without learning ${\mathbf{H}}_{\text{ima}, k}$, and ``no interference'' indicates the case with no multipath interference.}
\label{fig-result-interference}
\end{figure*}

This subsection evaluates the effectiveness of the proposed imaging algorithm under multipath interference in Sec.~\ref{subsec-enhanced-imager-multipath}.
It is assumed that ROI-irrelevant multipaths have been repetitively measured and removed from the CSI measurements $\widehat{\mathbf{H}}_{\text{sen},k}$, following the background calibration approach in Sec. \ref{subsubsec-channel-calibration}.
The ROI image is then reconstructed under the influence of the unknown multipath component $\mathbf{H}_{\text{roi-bg}}$ defined in \eqref{eq-joint-learning-channel}.
Based on \eqref{eq-joint-learning-channel} and \eqref{eq-calibration-channel-measurement}, the signal-to-interference ratio (SIR) for imaging is defined as
\begin{equation}
\text{SIR} = \frac{\|\mathbf{H}_{\text{tx-roi-rx}} + \mathbf{H}_{\text{tx-roi-ris-rx}, k} + \mathbf{H}_{\text{tx-ris-roi-rx}, k}\|_2^2}{\|\mathbf{H}_{\text{roi-bg}}\|_2^2}.
\end{equation}
Additionally, the interference recovery accuracy (IRA) is defined as
\begin{equation}
\text{IRA} = \frac{\|\widehat{\mathbf{A}}\widehat{\boldsymbol{\sigma}}-\mathbf{A}\boldsymbol{\sigma}\|_2^2}{\|\mathbf{A}\boldsymbol{\sigma}\|_2^2},
\end{equation}
where $\widehat{\mathbf{A}}$ and $\widehat{\boldsymbol{\sigma}}$ denote the estimates of $\mathbf{A}$ and $\boldsymbol{\sigma}$, respectively.
We consider $K=40$ and $D=40\lambda$, and compare the proposed joint imaging and interference learning method against conventional imaging without interference modeling.

As shown in Fig. \ref{fig-result-interference}, the proposed method significantly outperforms the baseline across various SIR levels. When the SIR exceeds 0 dB, the method can effectively optimize both the image representation $\mathcal{M}_{\boldsymbol{\theta}}$ and the multipath-related matrix $\mathbf{A}$, achieving imaging performance comparable to the interference-free case.
However, performance gradually degrades as SIR falls below 0 dB.
The matrix $\mathbf{A}$ becomes untrainable when SIR drops below -20 dB.
These results indicate that the proposed INR-based imager remains effective under a partially known forward model, provided that the interference power is lower than that of the modeled components.
In contrast, conventional imaging without interference learning is anticipated to perform well only when SIR exceeds 20 dB in the considered scenario, due to the mismatch between the assumed and actual multipath channels.
Furthermore, the results labeled ``compared'' can also reveal the impacts of calibration errors in Sec. \ref{subsubsec-channel-calibration}, which are not learned by the proposed method.

\subsubsection{Successive Imaging Performance for Dynamic Targets}

This subsection evaluates the performance of successive imaging for dynamic targets using the prior-embedding method described in Sec.~\ref{subsec-enhanced-imager-dynamic}.
We consider three successive time instants, during which two people move within the ROI.
Their body shapes change over time, as shown in the ``GT'' row of Table~\ref{tab-result-dynamic}.
However, the three images remain highly correlated, allowing the image at each subsequent instant to benefit from the prior representation.
At time instant $t=1$, we use 5,000 training epochs to obtain an accurate image with an SSIM of 0.8898.
At the following two instants, only 200 training epochs are allowed in order to accelerate the imaging process and increase the frame rate.
As shown in Table~\ref{tab-result-dynamic}, training from scratch at later instants results in distorted images and low quality.
In contrast, initializing the neural network with priors from time instant $t=1$ enables visually accurate reconstructions after just 200 epochs, with significantly higher PSNR and SSIM values compared to training from scratch.

\begin{figure}
    \centering
    \includegraphics[width=0.8\linewidth]{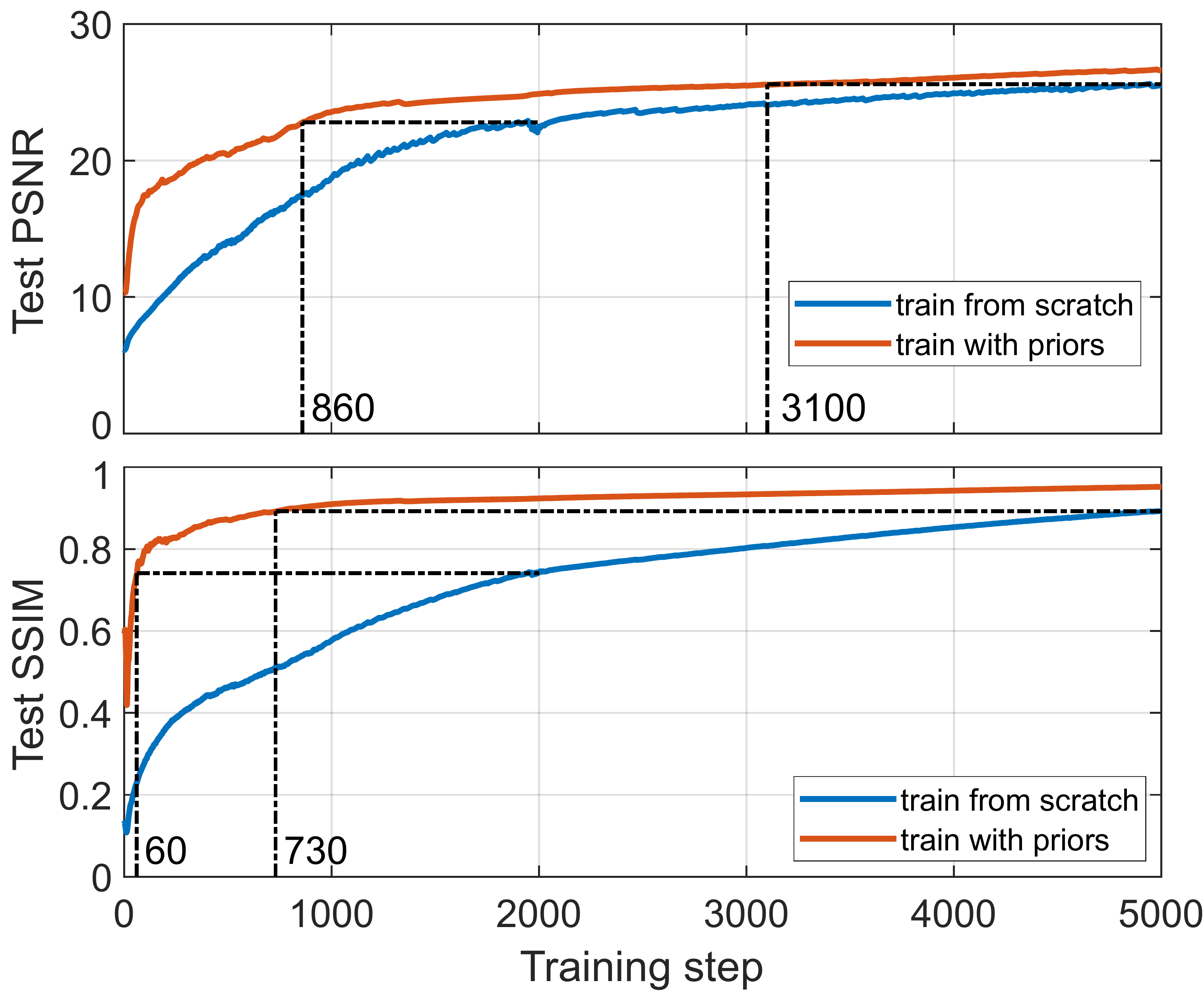}
    \captionsetup{font=footnotesize} 
    \caption{Test PSNR and SSIM during the training process, comparing models trained from scratch and with priors.}
    \label{fig-result-train-process-dynamic}
\end{figure}

Fig.~\ref{fig-result-train-process-dynamic} presents the test PSNR and SSIM values during training for the second time instant.
The results show that training with priors can achieve the PSNR performance of training from scratch at epoch 2,000 and 5,000 by epoch 860 and 3,100, respectively.
Similarly, the same SSIM achieved by training from scratch at epoch 2,000 (or 5,000) is reached with priors by epoch 60 (or 730).
Therefore, embedding implicit priors from the initial instant significantly improves imaging quality with fewer training epochs, enabling efficient successive imaging for dynamic targets.

\begin{table}[t]
    \renewcommand{\arraystretch}{1.4}
    \centering
    \fontsize{8}{8}\selectfont
    \captionsetup{font=small}
    \captionof{table}{Imaging results for dynamic targets.}\label{tab-result-dynamic}
    \begin{threeparttable}
        \begin{tabular}{c r ccc}
            \specialrule{1pt}{0pt}{-1pt}
            \xrowht{10pt} & Time instant $t$ & 1 & 2 & 3 \\
            & GT & 
            \adjustbox{valign=m}{\includegraphics[width=1cm]{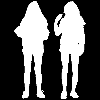}} & 
            \adjustbox{valign=m}{\includegraphics[width=1cm]{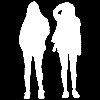}} & 
            \adjustbox{valign=m}{\includegraphics[width=1cm]{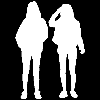}} \\
            \specialrule{0.5pt}{1pt}{1pt}
            \multirow{4}*{\vspace{-0.5cm}\makecell[c]{Train\\from\\scratch}} & Epoch & 5,000 & 200 & 200 \\
            & PSNR (dB) & 24.0932 & 10.2783 & 10.3673 \\
            & SSIM & \textbf{0.8898} & 0.3653 & 0.3726 \\
            & Result & 
            \adjustbox{valign=m}{\includegraphics[width=1cm]{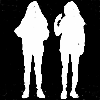}} & 
            \adjustbox{valign=m}{\includegraphics[width=1cm]{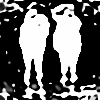}} & 
            \adjustbox{valign=m}{\includegraphics[width=1cm]{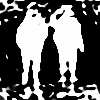}} \\
            \specialrule{0.5pt}{1pt}{1pt}
            \multirow{4}*{\vspace{-0.5cm}\makecell[c]{Train\\with\\priors}} & Epoch & / & 200 & 200 \\
            & PSNR (dB) & / & 18.4271 & 17.3824 \\
            & SSIM & / & \textbf{0.8171} & \textbf{0.8278} \\
            & Result & / & 
            \adjustbox{valign=m}{\includegraphics[width=1cm]{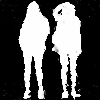}} & 
            \adjustbox{valign=m}{\includegraphics[width=1cm]{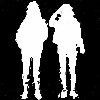}} \\
            \specialrule{1pt}{1pt}{0pt}
        \end{tabular}
    \end{threeparttable}
\end{table}

\begin{figure}
    \centering
    \includegraphics[width=0.8\linewidth]{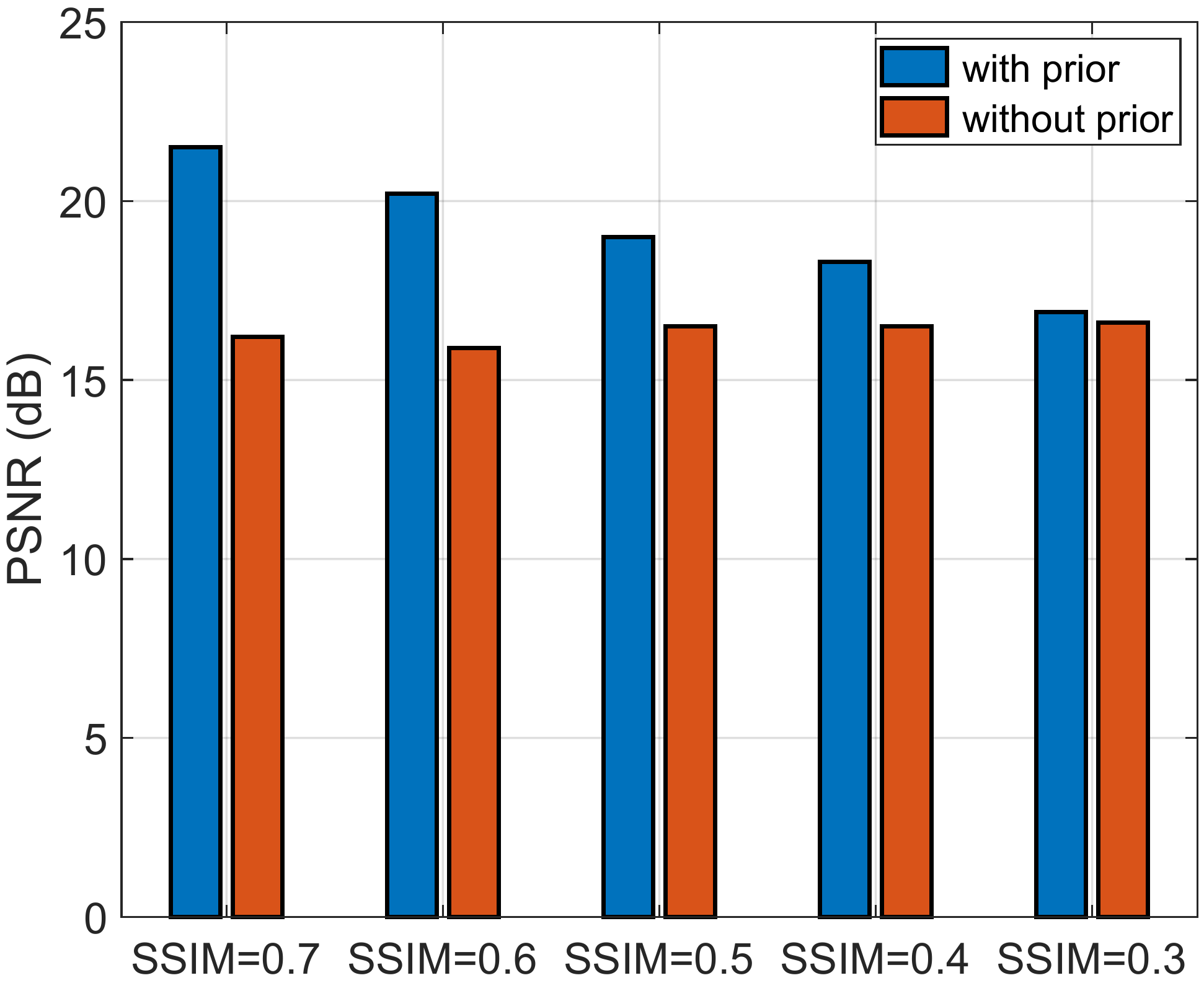}
    \captionsetup{font=footnotesize}
    \caption{PSNR performance with/without image priors for different SSIM values.}
    \label{fig-dynamic}
\end{figure}

Furthermore, we evaluate the prior-embedded performance gain across diverse image correlation levels.
Specifically, we employ the image at $t=1$ in Table \ref{tab-result-dynamic} as the prior image.
However, the subsequent target image is varied, and the SSIM metric is used to measure the image correlation between subsequent and prior images.
The simulation results at epoch 600, presented in Fig. \ref{fig-dynamic}, reveal that the performance gain originating from the prior-embedded NN parameters gradually degrades as image correlation decreases.
When the SSIM value between the prior and target images is reduced to 0.3, nearly the same PSNR performance is achieved regardless of whether prior information has been utilized.
However, when the number of training epochs is lower than 600, the proposed prior-aided imaging method consistently exhibits a rapid convergence rate, enabling coarse image formulation within a short duration, as depicted in Fig. \ref{fig-result-train-process-dynamic}.

\subsubsection{Imaging-Augmented Communication Performance} 

We evaluate the enhancement of communication performance using high-quality imaging results obtained from the INR-based method.
The transmit power is set to $P_{\text{t}}=0$ dBm, and we assume that the LOS path between TX and the user is not obstructed.
The TX employs $N_{\text{t}}=2$ antennas.
The RIS phase optimization algorithm proposed in \cite{wu2018intelligent} is adopted, and the RIS consists of $10\times10$ elements to reduce the computational complexity.
We respectively consider two scenarios where the user is either inside or outside the ROI.

\begin{figure}[t]
\centering
\captionsetup{font=footnotesize}
\begin{subfigure}[b]{0.48\linewidth}
\centering
\includegraphics[width=\linewidth]{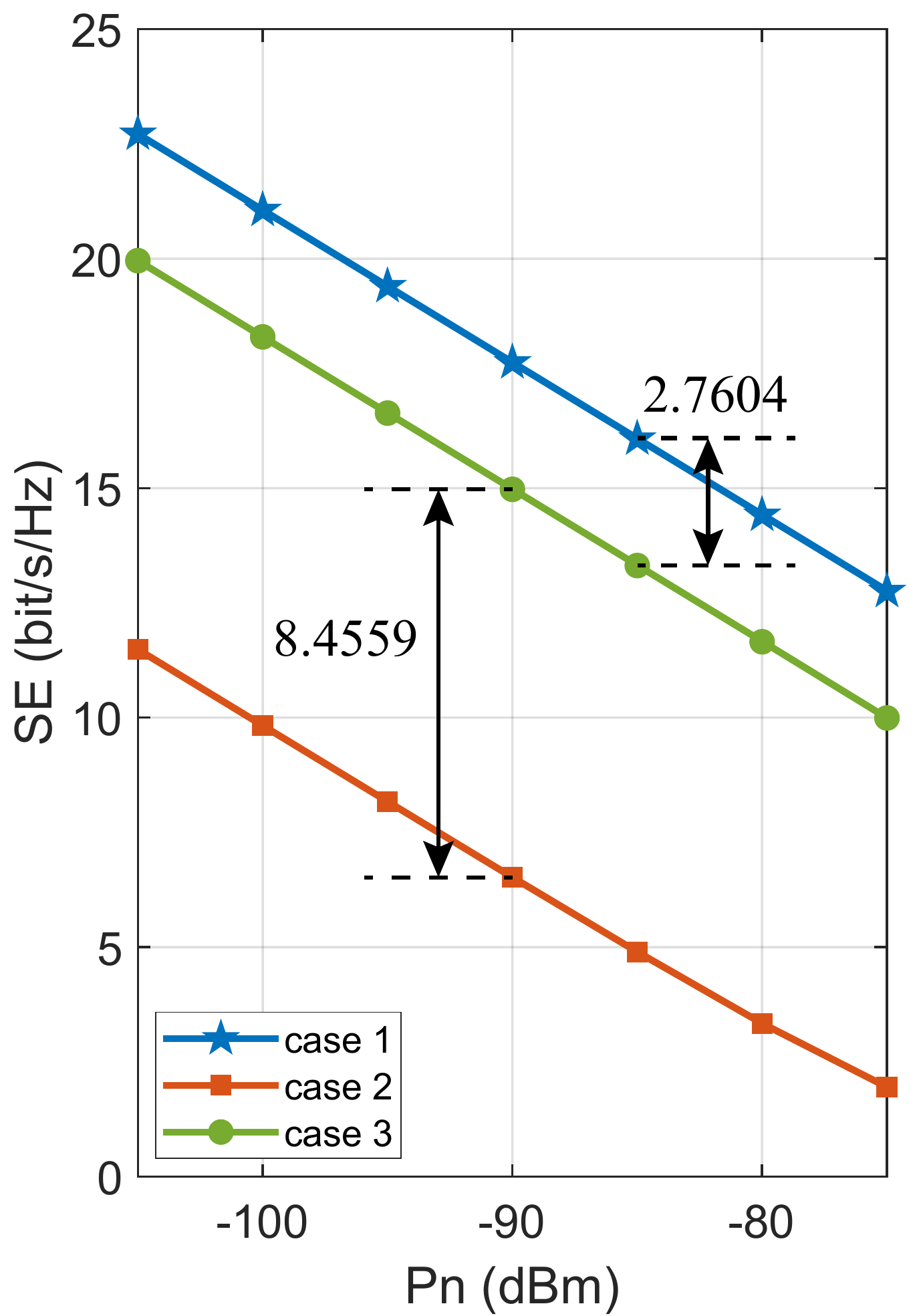}
\caption{User inside the ROI}
\label{fig-result-communication-inside}
\end{subfigure}
\begin{subfigure}[b]{0.48\linewidth}
\centering
\includegraphics[width=\linewidth]{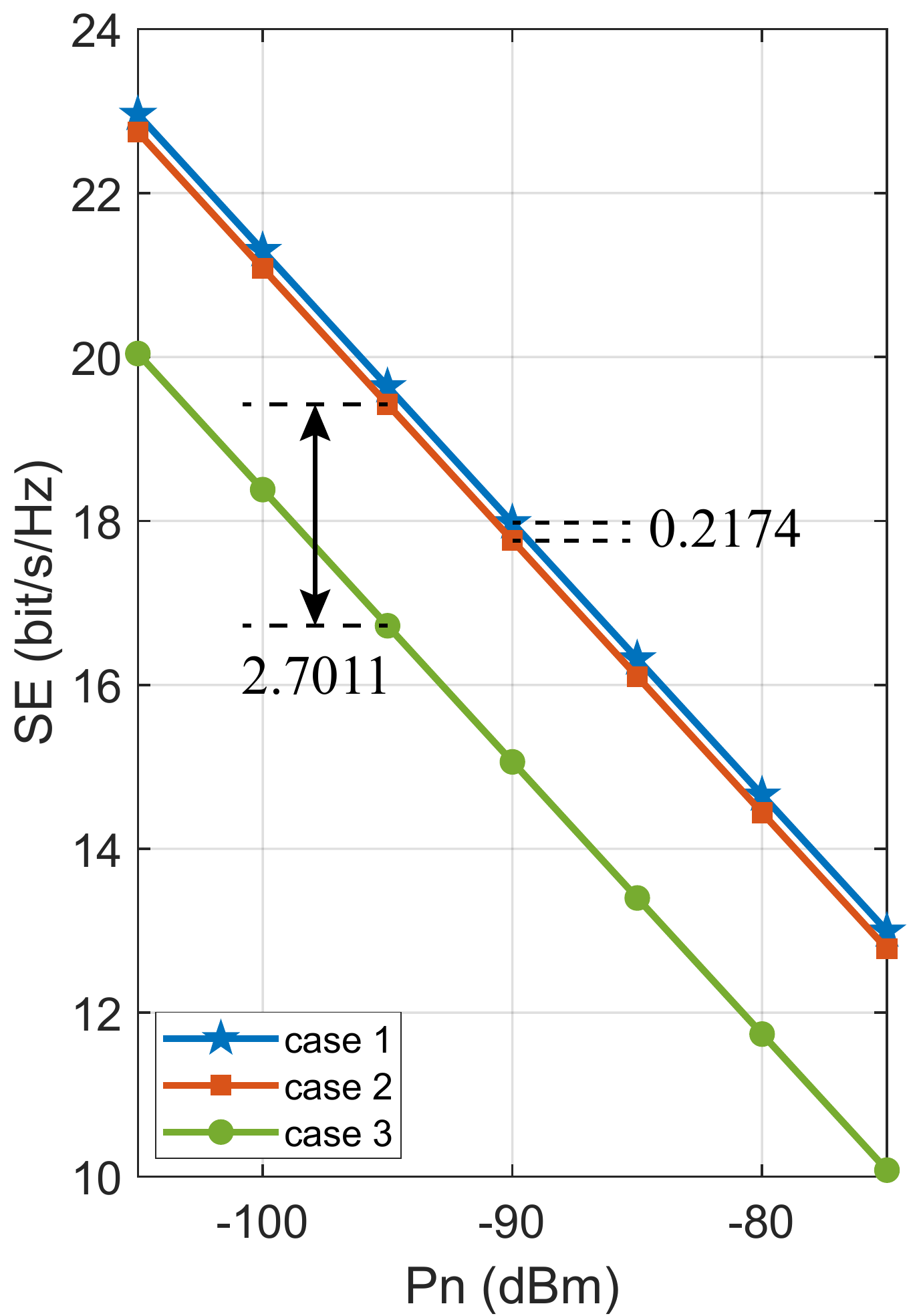}
\caption{User outside the ROI}
\label{fig-result-communication-outside}
\end{subfigure}

\caption{Communication SE versus noise power. ``case 1'' denotes optimized RIS phases with information provided by imaging results, ``case 2'' denotes optimized RIS phases without information provided by imaging results, and ``case 3'' denotes random RIS phases.}
\label{fig-result-communication}
\end{figure}

When the user is located inside the ROI, it simultaneously serves as both the communication terminal and the target for sensing, as described in Sec.~\ref{subsec-augment-commun-inside}.
We assume that the rough user location is $[40\lambda, 0, 0]^{\text{T}}$, which is used for RIS phase optimization without imaging results.
However, the actual position of the communication device (the user hand) is $[40\lambda, 10\lambda, 10\lambda]^{\text{T}}$, which can be identified using the INR-based high-quality image and computer vision techniques.
Fig.~\ref{fig-result-communication-inside} shows the SE  under random and optimized RIS phase configurations.
The results indicate that optimizing RIS phases based on inaccurate user positions may significantly degrade communication performance, potentially resulting in lower SE than with random RIS phase settings.
In contrast, optimizing the RIS phases using the accurate terminal location obtained from the imaging results increases the SE by 2.76 bit/s/Hz compared to the random configuration.

When the user is located outside the ROI, the targets within the ROI provide additional multipaths to the communication channel, as explained in Sec.~\ref{subsec-augment-commun-outside}.
The user is assumed to be located at $[30\lambda, -20\lambda, 20\lambda]^{\text{T}}$, which has been previously estimated for RIS phase optimization.
The corresponding SE results are illustrated in Fig. \ref{fig-result-communication-outside}.
It is observed that RIS phase optimization improves SE even when ROI-related multipaths are not considered in case 2, as their energy is typically low and may have limited influence on signal power.
However, incorporating environmental information obtained from the imaging results in case 1 leads to further improvement in SE, thereby enhancing the communication experience.
This performance gain is attainable only through the INR-based imager, which produces high-quality images, unlike traditional FT and CS methods that may not provide such information, as compared in Table \ref{tab-result-compare-ft-cs}.

\section{Conclusion and Future Research Directions}
\label{sec-conclusion}

This study presents a RIS-aided ISAC system in which the communication infrastructure is reused to enable target imaging.
The INR technique is introduced for wireless imaging, where the NN architecture is specifically designed using positional encoding and a sine activation function to embed image information within the NN parameters.
Under the supervision of physical models, DL techniques are employed to optimize the image representation, offering several advantages over traditional model-based and data-driven methods.
To improve imaging performance in practical environments, challenges such as multipath interference and dynamic target imaging are addressed.
Furthermore, the imaging results are utilized to enhance communication SE.
Simulation results demonstrate that the proposed algorithm outperforms state-of-the-art methods by achieving high imaging accuracy with low measurement overhead.
Multipath interference can be effectively learned to support high-quality imaging, and the use of priors facilitates imaging for dynamic targets.
Additionally, the communication performance of users located either inside or outside the ROI can be improved through the assistance of imaging results.

Our future work may include optimizing the NN with advanced positional encoding methods, activation functions, and accelerated training strategies.
Moreover, the INR-based imaging algorithm is anticipated to be applied to diverse scenarios, including imaging using multiple observation views and multiple frequency bands.
Validation through field trials conducted in live commercial systems is also required to verify the practical applicability of the proposed algorithms.

\begin{appendices}
\section{}\label{appendix-ris-phase-opt}

According to the channel models presented in Sec. \ref{sec-system-model} and Sec. \ref{sec-imaging-augmented-communication}, we have $\mathbf{h}_{\text{com}}(\boldsymbol{\omega})=\mathbf{B}\boldsymbol{\omega}+\mathbf{c}$, where $\mathbf{B} = g_{\text{com}}g_{\text{ris}}\mathbf{H}_{\text{tx-ris}}\text{diag}(\mathbf{h}_{\text{ris-ue}})$ and $\mathbf{c}=\mathbf{h}_{\text{tx-ue}}$ when the user is inside the ROI, while $\mathbf{B} = g_{\text{com}}g_{\text{ris}}(\mathbf{H}_{\text{tx-ris}} \text{diag}(\mathbf{h}_{1}) + \mathbf{H}_{1} \text{diag}(\mathbf{h}_{\text{ris-ue}}))$ and $\mathbf{c}=\mathbf{h}_{\text{tx-ue}} + \mathbf{h}_{\text{tx-roi-ue}}$ when the user is outside the ROI.
Consequently, the problem (P5) can be rewritten as
\begin{equation*}
\begin{aligned}
\text {(P6)}\quad & \max \limits_{\boldsymbol{\omega}} && \boldsymbol{\omega}^{\text{H}}\mathbf{B}^{\text{H}}\mathbf{B}\boldsymbol{\omega}+\boldsymbol{\omega}^{\text{H}}\mathbf{B}^{\text{H}}\mathbf{c}+\mathbf{c}^{\text{H}}\mathbf{B}\boldsymbol{\omega} \\
& \text { s.t. } && |\omega_{n_{\text{s}}}|=1, \ \ \forall n_{\text{s}}=1, \ldots, N_\text{s}. \\
\end{aligned}
\end{equation*}
By introducing an auxiliary variable $z$, problem (P6) is equivalently written as
\begin{equation*}
\begin{aligned}
\text {(P7)}\quad & \max \limits_{\bar{\boldsymbol{\omega}}} && \bar{\boldsymbol{\omega}}^{\text{H}}\mathbf{R}\bar{\boldsymbol{\omega}} \\
& \text { s.t. } && |\bar{\omega}_{n_{\text{s}}}|=1, \ \ \forall n_{\text{s}}=1, \ldots, N_\text{s}+1, \\
\end{aligned}
\end{equation*}
where
\begin{equation}
\mathbf{R}=\begin{bmatrix}
 \mathbf{B}^{\text{H}}\mathbf{B} & \mathbf{B}^{\text{H}}\mathbf{c}\\
 \mathbf{c}^{\text{H}}\mathbf{B} & \mathbf{0}
\end{bmatrix}, \quad \bar{\boldsymbol{\omega}}=\begin{bmatrix}
 \boldsymbol{\omega} \\
 z
\end{bmatrix}
\end{equation}
Define $\mathbf{W}=\bar{\boldsymbol{\omega}}\bar{\boldsymbol{\omega}}^{\text{H}}$, we have $\bar{\boldsymbol{\omega}}^{\text{H}}\mathbf{R}\bar{\boldsymbol{\omega}}=\text{tr}(\mathbf{R}\bar{\boldsymbol{\omega}}\bar{\boldsymbol{\omega}}^{\text{H}})=\text{tr}(\mathbf{RW})$, where $\mathbf{W}\succeq 0$ and $\text{rank}(\mathbf{W})=1$ should be satisfied.
Here, $\mathbf{W}\succeq 0$ means that the matrix $\mathbf{W}$ is positive semidefinite.
By applying semidefinite relaxation, problem (P7) is reduced to
\begin{equation*}
\begin{aligned}
\text {(P8)}\quad & \max \limits_{\mathbf{W}} && \text{tr}(\mathbf{RW}) \\
& \text { s.t. } && \mathbf{W}_{n_\text{s}, n_{\text{s}}}=1, \ \ \forall n_{\text{s}}=1, \ldots, N_\text{s}+1, \\
&&& \mathbf{W}\succeq 0. \\
\end{aligned}
\end{equation*}
Problem (P8) is a standard convex semidefinite program, and it can be optimally solved by convex optimization solvers such as CVX.
Furthermore, additional steps are presented in \cite{wu2018intelligent} to construct a rank-one solution from the optimal higher-rank solution of problem (P8).

\end{appendices}

\bibliographystyle{IEEEtran}
\bibliography{trans_ref}{}

\end{document}